\documentclass[11pt]{article}
\usepackage{mathrsfs}
\usepackage{amsmath}
\usepackage{amssymb}
\usepackage{amsfonts}
\usepackage{CJK}
\usepackage{subfigure}
\usepackage{graphicx}
\usepackage{dsfont}
\usepackage{amscd}
\usepackage[body={16.5cm,24cm}, top=1cm]{geometry}
\newtheorem{thm}{Theorem}

\begin{document}

\title{Darboux Transformation and Classification of Solution for Mixed Coupled Nonlinear Schr\"{o}dinger Equations}
\author{Liming Ling$^1$, Li-Chen Zhao$^2$, Boling Guo$^3$\\
$^1$School of Mathematics, South China University of Technology, Guangzhou 510640, China;\\ Email: linglm@scut.edu.cn\\
$^2$Department of Physics, Northwest University, 710069, Xi'an, China;\\
$^3$Institute of Applied Physics and Computational Mathematics, 100088, Beijing, China\\}

\maketitle
\begin{abstract}
We derive generalized nonlinear wave solution formula
for mixed coupled nonlinear Sch\"odinger equations¡¡(mCNLSE) by performing the unified Darboux transformation.
We give the classification of the general
soliton formula on the nonzero background based on the dynamical behavior.
Especially, the conditions for breather, dark soliton and rogue wave solution for mCNLSE are given in detail.
Moreover, we analysis the interaction between dark-dark soliton solution and breather solution.
These results would be helpful for nonlinear localized wave excitations and applications in vector nonlinear systems.
\


\end{abstract}

\maketitle

\section{Introduction}

Nonlinear Schr\"odinger equation (NLSE) is an important model in mathematical physics,
which can be applied to hydrodynamics \cite{Z-S}, plasma physics \cite{ions}, molecular biology \cite{K-M} and optics \cite{Agrawal}.
Recently, Peregrine soliton (rogue wave solution), Akhmediev breather, Kuznetzov-Ma breather and dark soliton were observed in experiments in succession.
For instance, Kuznetzov-Ma soliton was confirmed in 2012 \cite{K-M}, the Akhmediev breather was verified in numerical experiment \cite{Akhmediev},
the Peregrine soliton was experimentally observed in nonlinear fibre optics system \cite{Peregrine}, water tank \cite{Rogue1,Rogue2} and plasma \cite{plasma}.
Dark soliton was observed on the surface of water \cite{dark}.
Indeed those exact solutions for the NLSE on the plane wave background were  known well long time ago \cite{K-M}.
For the focusing NLSE, there exists Akhmediev breather, Kuznetzov-Ma soliton and Peregrine soliton.
There exists dark soliton for the defocusing NLSE.

However, the nonlinear localized wave solutions for the coupled nonlinear Schr\"odinger equations (CNLSE) are more complexity than NLSE.
Firstly, the spectral problem of
NLSE is $2\times 2$, the coupled
NLSE is $3\times 3$. The inverse scattering method of CNLSE on the nonzero background has not been solved completely \cite{Ablowitz}. Secondly,
for the Darboux transformation (DT) method, the Darboux matrix for $3\times 3$ spectral problem is more
complexity than $2\times 2$ spectral problem. What's more, the Darboux matrix for the defocusing or mixed coupled NLSE
is no longer positive or negative definite. In this work, we will use the matrix analysis method to deal with the condition of
positive or negative definite of Darboux matrix.

Previous to introducing our work, we review a brief research history of CNLSE.
The integrability of the CNLSE had been
shown by Manakov who had also obtained the bright soliton
in a focusing medium by applying the inverse scattering
method \cite{Manakov}. An interesting fact for CNLSE is the collision of soliton on the vanishing background
can be inelastic \cite{Lakshmanan1,lakshmanan}
or even appear the soliton reflection \cite{Wang-Zhang}.
For the nonzero background, there are lots of works for the focusing CNLSE \cite{Wright,Guo,Degasperis,zhao,zhao1,wang-chen,He1}.
But there are few results for the defocusing and mixed case \cite{Ablowitz}. The dark-dark soliton solution, bright-dark soliton solution and breather solution
for defocusing CNLSE were given in reference \cite{Prinari} by inverse scattering method.
The soliton solutions for the multi-component NLSE were given in reference \cite{Wright,Park,Tsuchida}
through DT. Recently,
different types of soliton solutions on the nonzero background were obtained by algebraic geometry reduction method \cite{Kalla}.

DT is a powerful method to construct the
soliton solution for the integrable equations. There are different methods to
derive the DT, for instance, operator decomposition method \cite{Deift}, gauge transformation \cite{Matveev,Gu}, loop group method \cite{TU} and Riemann-Hilbert
method \cite{NMZP}.
The DT for multi-component NLSE was given in reference \cite{Wright,Park,Wright1,Degasperis1,Degasperis2} (and reference therein).
Recently, combined Darboux dressing and
tau function method for dark-dark soliton is given in reference \cite{Blas}, Tsuchida gave a detail analysis of different types of solutions by
using the Darboux-B\"acklund transformation \cite{Tsuchida}.

In this work, we focus on the nonlinear localized wave solutions of the mCNLSE
\begin{equation}\label{mvnls}
    \begin{split}
      {\rm i} q_{1,t}+\frac{1}{2}q_{1,xx}+(|q_2|^2-|q_1|^2)q_1 &=0,  \\
      {\rm i} q_{2,t}+\frac{1}{2}q_{2,xx}+(|q_2|^2-|q_1|^2)q_2 &=0,
    \end{split}
\end{equation}
on the plane wave background.
The classification of
nonlinear localized wave solutions for mCNLSE \eqref{mvnls} on the nonzero background is not an easy work but be important for nonlinear wave theory and physical applications. Indeed,
there are few results about the classification of nonlinear localized wave solutions for CNLSE, even for the focusing CNLSE.
Recently, there are lot of works about focusing CNLSE through DT.
For instance, vector rogue wave (type-I and type-II) solution,
bright-dark-rogue wave solution \cite{Guo,Degasperis,zhao,zhao1} and high order solution \cite{Ling1} have been given.
However, for the mCNLSE or the defocusing CNLSE,
the Darboux matrix is no longer always positive definite or negative definite. We would confront
 with a new obstacle for the classification of solution through DT.
Comparing with focusing CNLSE, a key step is giving the positive or negative definite condition for the Darboux matrix.
To the best of our knowledge, this problem have never been solved with a proper method.
In this work, we use the matrix analysis method to deal with this problem. Through this method, we can obtain
the complete classification for the nonsingular solution for the mCNLSE \eqref{mvnls}. Besides the DT,
the inverse scattering method \cite{Wang-Zhang} and the bilinear method \cite{Lakshmanan1,feng,zhangdajun}
can be used to derive soliton solution of the CNLSE too.

This paper is organized as following. In section 2, we classify the soliton solution of mCNLSE \eqref{mvnls} with two categories.
Denote the wave vector of plane wave background for the $i$-th component is $a_i$ respectively.
The first case is $a_1=a_2$.  In this case, we can obtain the degenerate rogue wave, degenerate breather (type-I and type-II),
degenerate dark-dark soliton and bright-dark soliton.
The second case
is $a_1\neq a_2$. In this case, we can obtain the general rogue wave solution, general breather (type I and type II) solution and general dark-dark soliton solution.
The coexistence of different types solution is given by two different categories. The rogue wave solutions for mCNLSE \eqref{mvnls} are given with a compact
form. The classification of rogue wave solutions is given based on the dynamics behavior. A method for looking for different types of
rogue wave is given in detail.  In section 3, we give the interaction
between different types of soliton solution. Since the
rogue wave solution is the limit of breather solution,
we deem that rogue wave is the same kind with breather solution, although their dynamics behaviors are different.
Thus we merely analysis the interaction between breather type solution and dark soliton. Finally, we give some discussions and conclusions.

\section{Darboux transformation and classification of solution}
As we well known that, the mCNLSE admits the following Lax pair:
\begin{subequations}\label{lax}
  \begin{eqnarray}
    \Phi_x &=&U\Phi,\,\,U(\lambda,Q)\equiv {\rm i}(\lambda \sigma_3+Q), \\
    \Phi_t &=&V\Phi,\,\, V(\lambda,Q)\equiv {\rm i}\lambda^2\sigma_3+{\rm i}\lambda Q-\frac{1}{2}\sigma_3({\rm i}Q^2-Q_x),
  \end{eqnarray}
\end{subequations}
where
\begin{equation*}
  \sigma_3=\mathrm{diag}(1,-1,-1), \,\,\, Q=\begin{bmatrix}
                                            0 & -\bar{q}_1 & \bar{q}_2 \\
                                            q_1 & 0 & 0 \\
                                            q_2 & 0 & 0 \\
                                          \end{bmatrix},
\end{equation*}
the overbar represent the complex conjugation (similarly hereinafter). The compatibility condition of Lax pair \eqref{lax} gives the mCNLSE \eqref{mvnls}.
The unified DT is obtained in reference \cite{Ling} with an integral form. Here we give another representation
with a limit form:
\begin{thm}[\cite{Ling}, Ling, Zhao and Guo]
The following unified DT
\begin{equation}\label{edt}
 \Phi[1]=T_1\Phi,\,\, T_1=I-\frac{P_1}{\lambda-\bar{\lambda}_1},
\end{equation}
where
\begin{subequations}\label{edtpmatrix}
\begin{eqnarray}
\text{if  }\lambda_1\not\in\mathbb{R},\,\, P_1&=&\frac{(\lambda_1-\bar{\lambda}_1)|y_1\rangle\langle y_1|J}{\langle y_1|J|y_1\rangle},\label{edtpmatrixa}\\
\text{if  }\lambda_1\in\mathbb{R},\,\, P_1&=&\lim_{\lambda_1\rightarrow\bar{\lambda}_1}\frac{(\lambda_1-\bar{\lambda}_1)|y_1\rangle\langle y_1|J}{\langle y_1|J|y_1\rangle},
\label{edtpmatrixb}\end{eqnarray}
\end{subequations}
 $J=\mathrm{diag}(1,-1,1),$ $|y_1\rangle\equiv v_1(x,t)\Phi_1$, $\langle y_1|=|y_1\rangle^{\dag}$, $v_1(x,t)$ is an arbitrarily complex function, $\Phi_1$ is
 a special vector solution for linear system \eqref{lax} with $\lambda=\lambda_1$,
and $^{\dag}$ represents Hermite conjugation (similarly hereinafter),
converts the above linear system \eqref{lax} into a new linear system
\begin{subequations}\label{newlax}
  \begin{eqnarray}
    \Phi[1]_x &=&U[1]\Phi[1],\,\,U[1]=U(\lambda,Q[1]), \\
    \Phi[1]_t &=&V[1]\Phi[1],\,\,V[1]=V(\lambda,Q[1]),
  \end{eqnarray}
\end{subequations}
and transformation between potential functions is
\begin{equation}\label{potential}
    Q[1]=Q+[\sigma_3,P_1],
\end{equation}
where commutator $[A,B]\equiv AB-BA.$
\label{thm1}\end{thm}
 To keep the results with the inverse scattering method, we restrict the parameter
$\lambda_1\in \{z|\mathrm{Im}(z)\geq0\}.$ The expression for $P_1$ \eqref{edtpmatrixb}
is considered in the limited sense, since there exists special solutions which satisfy $\langle y_1|J|y_1\rangle=0$ when $\lambda_1\in\mathbb{R}$.

To obtain the nonlinear localized wave solutions on the plane wave background, we consider the following plane wave solutions as the seed solutions for DT
\begin{equation}\label{seed}
  \begin{split}
    q_1[0]&=c_1e^{{\rm i}\theta_1},\,\,\theta_1=\left[a_1x-\left(\frac{1}{2}a_1^2+c_1^2-c_2^2\right)t\right], \\
    q_2[0]&=c_2e^{{\rm i}\theta_2},\,\,\theta_2=\left[a_2x-\left(\frac{1}{2}a_2^2+c_1^2-c_2^2\right)t\right],
  \end{split}
\end{equation}
where $a_1$, $a_2$, $c_1$ and $c_2$ are real constants.
Through above transformation \eqref{potential} and the plane wave seed solution, we can obtain different types of nonlinear wave solutions.
If $\lambda_1\not\in \mathbb{R}$, the DT can be used to derive breather solution, rogue wave solution and bright-dark soliton solution.
If $\lambda_1\in \mathbb{R}$, the dark-dark soliton solution can be obtained through this transformation \eqref{potential}.

Next we give a method how to choose the special solution $\Phi_1$ to construct the nonsingular exact solution of mCNLSE \eqref{mvnls}.
Through above DT \eqref{edt} and \eqref{potential}, we can obtain many new nonlinear wave solutions for mCNLSE\eqref{mvnls}, which have not been reported before.
To use the transformation \eqref{edt}, we firstly need to solve the linear system \eqref{lax} with seed solution \eqref{seed} by
the gauge transformation method \cite{Guo}:
\begin{equation}\label{fund}
  \Phi=D\Psi,\,\,D=\mathrm{diag}(1,e^{{\rm i}\theta_1},e^{{\rm i}\theta_2}).
\end{equation}
Then matrix $\Psi$ satisfies the following linear system:
\begin{equation*}
\begin{split}
   \Psi_x&={\rm i}U_0\Psi,  \\
   \Psi_t&={\rm i}\left(\frac{1}{2}U_0^2+\lambda U_0-\frac{1}{2}\lambda^2+c_1^2-c_2^2\right)\Psi,
\end{split}
\end{equation*}
where
\begin{equation*}
    U_0=\begin{bmatrix}
          \lambda & -c_1 & c_2 \\
          c_1 & -\lambda-a_1 & 0 \\
          c_2 & 0 & -\lambda-a_2 \\
        \end{bmatrix}.
\end{equation*}

We can obtain the different kinds of solution by choosing different special solutions \eqref{fund}. The studies on coupled focusing or defocusing NLS have shown that the relative wave vector plays important role in determining dynamics of nonlinear waves \cite{zhao, zhao1, Degas}. Therefore, we classify them into two main cases according to the relative wave vector.

\subsection{Case I: When $a_2=a_1$ }
In this case, we have the following matrix decomposition
\begin{equation}\label{decom}
    U_0M=MD_0,\,\, D_0=\mathrm{diag}(\chi-\lambda,\mu-\lambda,-(\lambda+a_1)),\,\, \chi\neq\mu,
\end{equation}
where
\begin{equation*}
  M=\begin{bmatrix}
    1 & 1 & 0 \\[8pt]
    \frac{c_1}{\chi+a_1} & \frac{c_1}{\mu+a_1} & c_2 \\[8pt]
    \frac{c_2}{\chi+a_1} & \frac{c_2}{\mu+a_1} & c_1 \\
  \end{bmatrix},
\end{equation*}
and $\chi$ and $\mu$ are two different roots
of the quadratic equation:
\begin{equation}\label{quatic}
    \xi^2+(a_1-2\lambda)\xi-2a_1\lambda+c_1^2-c_2^2=0.
\end{equation}
Then the fundamental solution \eqref{fund} can be given as $\Phi=DMN$, where $N=\mathrm{diag}(e^{{\rm i} A_1},e^{{\rm i} B_1},e^{{\rm i} C_1})$,
\begin{equation*}
    \begin{split}
      A_1&=(\chi-\lambda)x+\left[\frac{1}{2}\chi^2+(c_1^2-c_2^2-\lambda^2)\right]t, \\
      B_1&=(\mu-\lambda)x+\left[\frac{1}{2}\mu^2+(c_1^2-c_2^2-\lambda^2)\right]t,  \\
      C_1&=-(a_1+\lambda)x+\left[\frac{1}{2}a_1^2+(c_1^2-c_2^2-\lambda^2)\right]t.
    \end{split}
\end{equation*}
By the decomposition \eqref{decom}, we can simplify the solution form of mCNLSE \eqref{mvnls} through using the following identity:
\begin{equation}\label{imp-eq}
    \begin{split}
      2\lambda-\chi+\frac{c_2^2-c_1^2}{\chi+a_1}=0,   \\
      2\lambda-\mu+\frac{c_2^2-c_1^2}{\mu+a_1}=0.
    \end{split}
\end{equation}
Interestingly,  a further classification on nonlinear wave solutions can  be made through relations between $c_1$ and $c_2$.

(a): When $c_1>c_2$, the system \eqref{mvnls} reflects the defocusing mechanism. With this case, there exists dark-dark soliton solution. Besides the dark-dark soliton, there exists breather solution. In what following, we give the explicit construction method for them. Firstly, solving equation \eqref{quatic}, we have the following
solution:
\begin{equation*}
  \begin{split}
    \chi= & \lambda-\frac{a_1}{2}+\sqrt{(\lambda+\frac{a_1}{2})^2-(c_1^2-c_2^2)}, \\
    \mu= & \lambda-\frac{a_1}{2}-\sqrt{(\lambda+\frac{a_1}{2})^2-(c_1^2-c_2^2)}.
  \end{split}
\end{equation*}
By above two roots, it is readily to obtain that $\mu+a_1=\frac{(\bar{\chi}+a_1)(c_1^2-c_2^2)}{|\chi+a_1|^2}$, which implies that the roots $\chi$ and $\mu$ can not lay on the upper half plane or lower half plane simultaneously.

To give the nonsingular solution of \eqref{mvnls}, one must choose the special solution $\Phi_1$ such that
$\Phi_1^{\dag}J\Phi_1$ is negative or positive definitely. To find the special solution $\Phi_1$, it is meaningful to analysis the following matrix:
\begin{equation*}
  M^{\dag}J M=\begin{bmatrix}
                    \frac{2(\bar{\lambda}-\lambda)}{\bar{\chi}-\chi}  & \frac{2(\bar{\lambda}-\lambda)}{\bar{\chi}-\mu} & 0 \\[8pt]
                    \frac{2(\bar{\lambda}-\lambda)}{\bar{\mu}-\chi}  & \frac{2(\bar{\lambda}-\lambda)}{\bar{\mu}-\mu} & 0 \\[8pt]
                     0 & 0 & c_1^2-c_2^2 \\
                   \end{bmatrix},
\end{equation*}
which can be obtained through equations \eqref{imp-eq}.
Indeed, the above matrix can not be positive definite or negative definite, since above matrix is nothing but the congruent matrix of $J$ and
the congruent matrix can not change the sign of characteristic roots. However, we can look for some submatrix which could be positive or negative definite.
By inspection,  if $\mathrm{Im}(\lambda)\mathrm{Im}(\chi)>0$, we can find that
\begin{equation*}
\begin{bmatrix}
  1 & \frac{c_1}{\bar{\chi}+a_1} & \frac{c_2}{\bar{\chi}+a_1} \\
  0 & c_2 & c_1 \\
\end{bmatrix}J\begin{bmatrix}
                     1 & 0 \\
                     \frac{c_1}{\chi+a_1}  & c_2 \\
                     \frac{c_2}{\chi+a_1} & c_1 \\
                   \end{bmatrix}=\begin{bmatrix}
                                   2\frac{\bar{\lambda}-\lambda}{\bar{\chi}-\chi} & 0 \\
                                   0 & c_1^2-c_2^2 \\
                                 \end{bmatrix}
\end{equation*}
is positive definite. Thus, if we choose the special solution $\Phi_1$ and $v_1(x,t)$ such that
\begin{equation}\label{ycase1}
    |y_1\rangle=D\begin{bmatrix}
e^{{\rm i}(\chi_1x+\frac{1}{2}\chi_1^2t)} \\
\frac{c_1}{\chi_1+a_1} e^{{\rm i}(\chi_1x+\frac{1}{2}\chi_1^2t)}+c_2\alpha_1e^{{\rm i}(-a_1x+\frac{1}{2}a_1^2t)} \\
\frac{c_2}{\chi_1+a_1}  e^{{\rm i}(\chi_1x+\frac{1}{2}\chi_1^2t)}+c_1\alpha_1e^{{\rm i}(-a_1x+\frac{1}{2}a_1^2t)} \\
 \end{bmatrix},
\end{equation}
where $\alpha_1$ is a nonzero complex constant.
Setting parameter
\begin{equation*}
    \alpha_1=\left(\frac{2\mathrm{Im}(\lambda_1)}{(c_1^2-c_2^2)\mathrm{Im}(\chi_1)}\right)^{1/2}e^{\frac{\beta_1}{2}+{\rm i}\gamma_1},
\end{equation*}
where $\beta_1$ and $\gamma_1$ are real constants, $\mathrm{Im}(\cdot)$ represents the image part of complex number $\cdot$ (similarly hereinafter),
then we can obtain the following breather solution by \textbf{theorem \ref{thm1}}
\begin{equation*}
    \begin{split}
        q_1[1] &=c_1\left[\frac{C_1+1}{2}+
        \frac{C_1-1}{2}\tanh(\frac{A_1-\beta_1}{2})+\frac{c_2D_1}{c_1}\mathrm{sech}(\frac{A_1-\beta_1}{2})e^{{\rm i}B_1}\right]e^{{\rm i}\theta_1},  \\
        q_2[1] &=c_2\left[\frac{C_1+1}{2}+
        \frac{C_1-1}{2}\tanh(\frac{A_1-\beta_1}{2})+\frac{c_1D_1}{c_2}\mathrm{sech}(\frac{A_1-\beta_1}{2})e^{{\rm i}B_1}\right]e^{{\rm i}\theta_1},
    \end{split}
\end{equation*}
where
\begin{equation*}
    \begin{split}
      A_1= & -2\mathrm{Im}(\chi_1)[x+\mathrm{Re}(\chi_1)t], \\
      B_1= & -(2\mathrm{Re}(\chi_1)+a_1)x+\left[\frac{1}{2}a_1^2-\mathrm{Re}(\chi_1^2)\right]t+\gamma_1+\frac{3}{2}\pi, \\
      C_1= & \frac{\bar{\chi}_1+a_1}{\chi_1+a_1},\,\, D_1= \left(\frac{
        2\mathrm{Im}(\lambda_1)\mathrm{Im}(\chi_1)}{c_1^2-c_2^2}\right)^{1/2},
    \end{split}
\end{equation*}
$\mathrm{Re}(\cdot)$ represents the real part of complex number $\cdot$ (similarly hereinafter).

\begin{figure}[htb]
\centering
\subfigure[$|q_1|^2$]{\includegraphics[height=50mm,width=80mm]{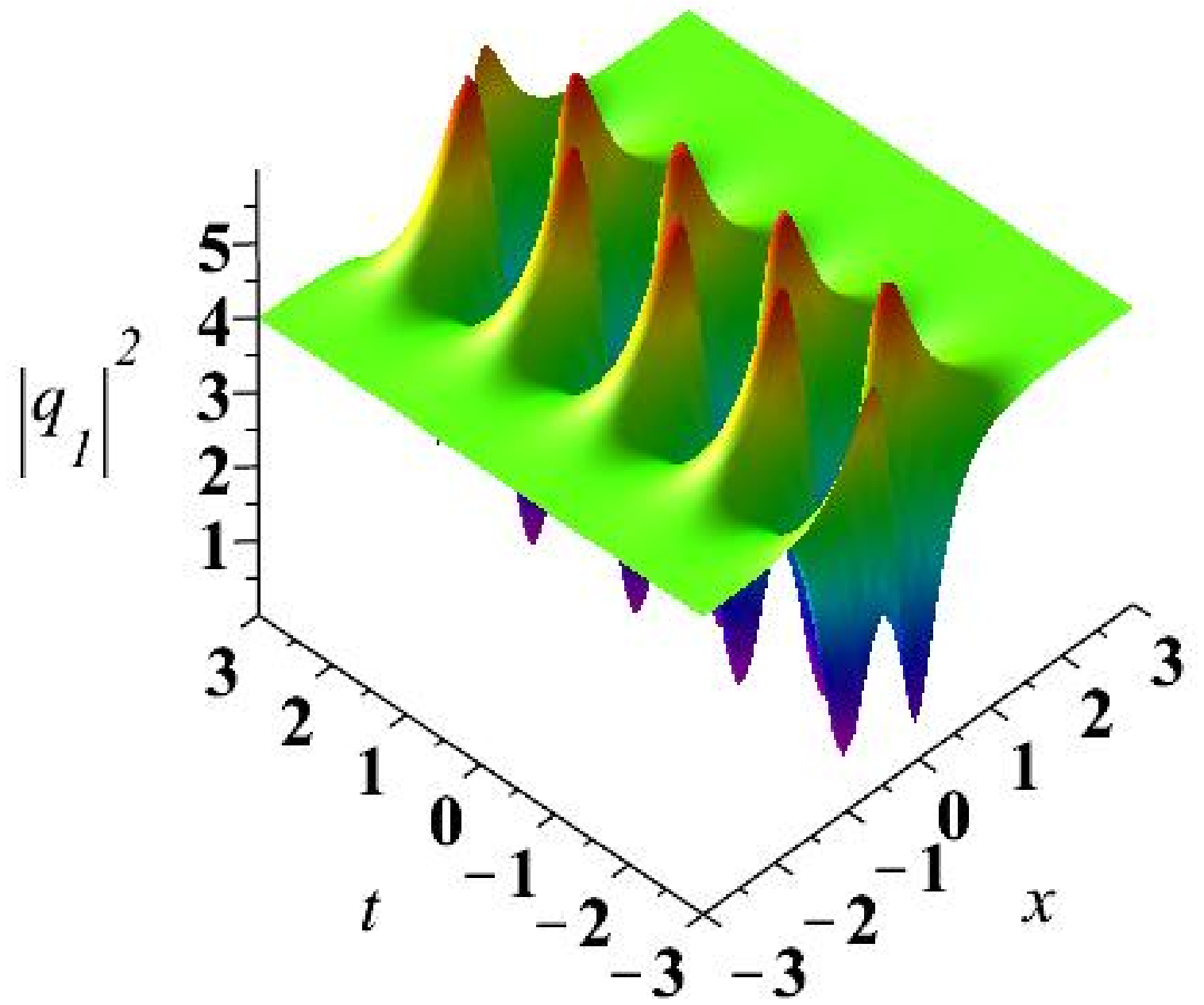}}
\hfil
\subfigure[$|q_2|^2$]{\includegraphics[height=50mm,width=80mm]{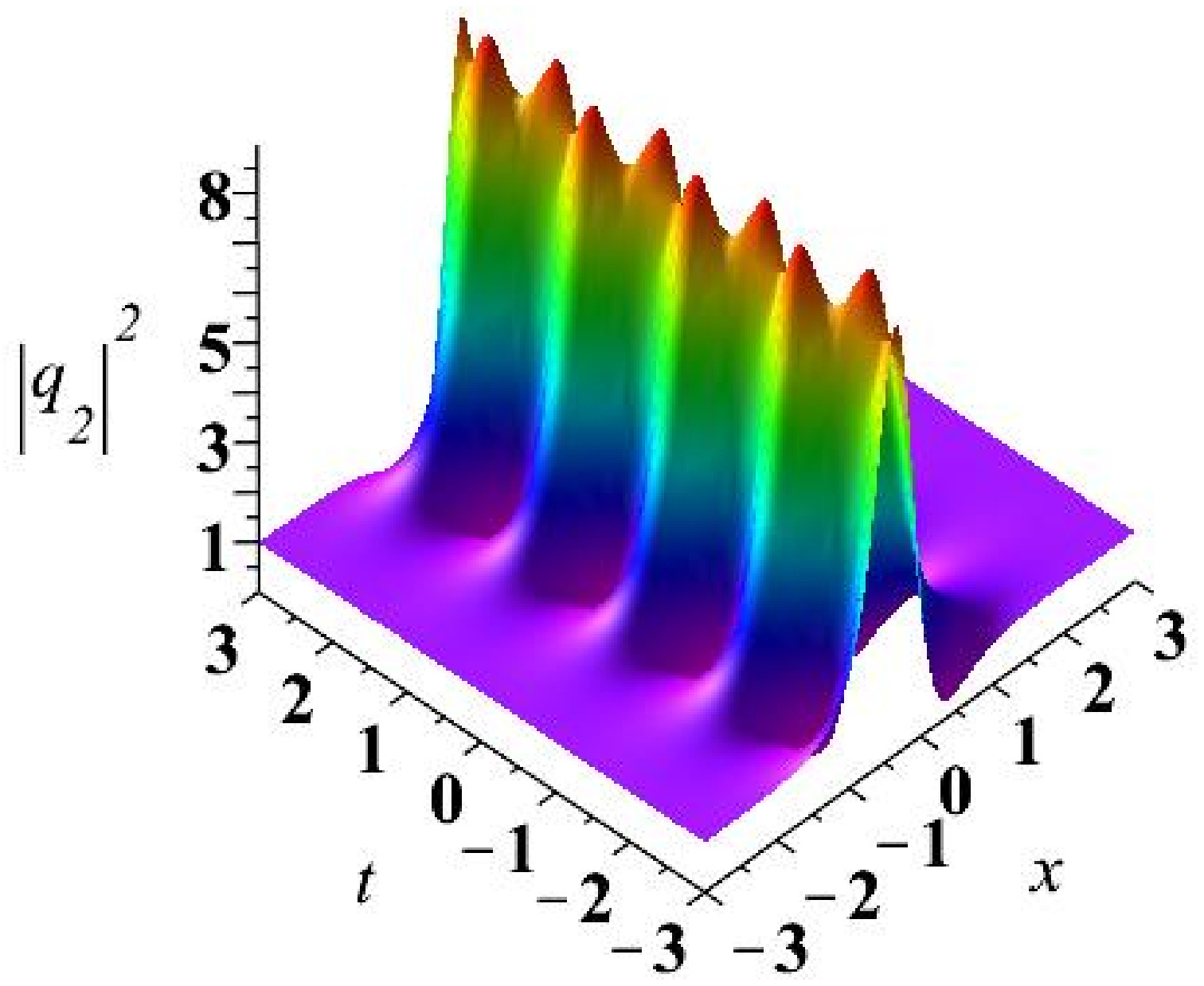}}
\caption{(color online): Breather-II type solution: Parameters $a_1=a_2=0$, $c_1=2$, $c_2=1$, $\lambda_1={\rm i}$, $\chi_1=
3{\rm i}$. It is seen that $|q_2|^2$ possesses the feature of bright soliton and breather.}\label{fig1}
\end{figure}

Through above expression of solution, we can see that this kind of breather solution is composed of a dark soliton solution and a bright soliton solution.
Notably, this type breather solution can not be derived to rogue wave solution through the limit method. This breather solution never appear in the single component NLSE. Thus we call this type of breather as breather-II to distinguish the breather which can be reduced to rogue wave. By choosing special parameters,
we can obtain the figure of breather-II (Fig. \ref{fig1}).

If we take $c_2=0$, we can obtain so-called bright-dark soliton solution
\begin{equation*}
    \begin{split}
       q_1[1] &=c_1\left[\frac{C_1+1}{2}+
        \frac{C_1-1}{2}\tanh(\frac{A_1-\beta_1}{2})\right]e^{{\rm i}\theta_1},  \\
        q_2[1] &=D_1\mathrm{sech}(\frac{A_1-\beta_1}{2})e^{{\rm i}(B_1+\theta_1)},
    \end{split}
\end{equation*}
where
\begin{equation*}
   \begin{split}
    A_1= & -2\mathrm{Im}(\chi_1)[x+\mathrm{Re}(\chi_1)t], \\
      B_1= & -(2\mathrm{Re}(\chi_1)+a_1)x+\left[\frac{1}{2}a_1^2-\mathrm{Re}(\chi_1^2)\right]t+\gamma_1+\frac{3}{2}\pi, \\
     C_1=&\frac{\bar{\chi}_1+a_1}{\chi_1+a_1}, \,\,
     D_1=\sqrt{2\mathrm{Im}(\lambda_1)\mathrm{Im}(\chi_1)}.
   \end{split}
\end{equation*}

(b): When $c_1=c_2$, there is no nontrivial solution of mCNLSE could be constructed through above \textbf{theorem} \ref{thm1}.

(c): When $c_1<c_2$, the system \eqref{mvnls} reflects the focusing mechanism.
By equation \eqref{imp-eq}, we have the following matrix
\begin{equation*}
    \begin{bmatrix}
  1 & \frac{c_1}{\bar{\chi}+a_1} & \frac{c_2}{\bar{\chi}+a_1} \\
   1 & \frac{c_1}{\bar{\mu}+a_1} & \frac{c_2}{\bar{\mu}+a_1} \\
\end{bmatrix}J\begin{bmatrix}
                     1 & 1 \\
                     \frac{c_1}{\chi+a_1}  &  \frac{c_1}{\mu+a_1}\\
                     \frac{c_2}{\chi+a_1} & \frac{c_2}{\mu+a_1} \\
                   \end{bmatrix}=\begin{bmatrix}
                                  \frac{2(\bar{\lambda}-\lambda)}{\bar{\chi}-\chi}  & \frac{2(\bar{\lambda}-\lambda)}{\bar{\chi}-\mu} \\[8pt]
                    \frac{2(\bar{\lambda}-\lambda)}{\bar{\mu}-\chi}  & \frac{2(\bar{\lambda}-\lambda)}{\bar{\mu}-\mu}
                                 \end{bmatrix}
\end{equation*}
is positive definite, which could be used to construct
the nonsingular solution.
If we choose the special solution $\Phi_1$ and $v_1(x,t)$ such that
\begin{equation}\label{ycase2}
    |y_1\rangle=D\begin{bmatrix}
               e^{{\rm i}(\chi_1x+\frac{1}{2}\chi_1^2t)}+\alpha_1 e^{{\rm i}(\mu_1x+\frac{1}{2}\mu_1^2t)} \\
                   \frac{c_1}{\chi_1+a_1} e^{{\rm i}(\chi_1x+\frac{1}{2}\chi_1^2t)}+\frac{c_1\alpha_1}{\mu_1+a_1} e^{{\rm i}(\mu_1x+\frac{1}{2}\mu_1^2t)} \\
   \frac{c_2}{\chi_1+a_1}  e^{{\rm i}(\chi_1x+\frac{1}{2}\chi_1^2t)}+\frac{c_2\alpha_1 }{\mu_1+a_1} e^{{\rm i}(\mu_1x+\frac{1}{2}\mu_1^2t)} \\
 \end{bmatrix},
\end{equation}
where $\alpha_1$ is a complex parameter.
Moreover, we introduce the following notation:
\begin{equation*}
    \lambda_1+\frac{a_1}{2}=\frac{\alpha}{2}\left(\zeta_1-\zeta_1^{-1}\right),\,\,\alpha=\sqrt{c_2^2-c_1^2},
\end{equation*}
it follows that
\begin{equation*}
\begin{split}
                    \sqrt{(\lambda_1+\frac{a_1}{2})^2+\alpha^2} &=\frac{\alpha}{2}\left(\zeta_1+\zeta_1^{-1}\right),  \\
                   \chi_1  &= \alpha\zeta_1-a_1,\\
                   \mu_1  &= -\alpha\zeta_1^{-1}-a_1.
\end{split}
\end{equation*}
Setting
\begin{equation*}
    \alpha_1=\exp[\alpha(1-|\zeta_1^{-2}|)\zeta_{1i}e_1-{\rm i}\alpha(1+|\zeta_1|^2)\zeta_{1r}f_1],\,\,\zeta_1=e^{\gamma_1+{\rm i}\beta_1},
\end{equation*}
where $\zeta_{1r}=\mathrm{Re}(\zeta_1)$, $\zeta_{1i}=\mathrm{Im}(\zeta_1),$
$e_1$ and $f_1$ are real constants, then we can obtain
\begin{equation*}
    q_1[1]=c_1\left[1+(e^{{\rm i}\beta_1}-e^{-{\rm i}\beta_1})\frac{\sinh(C+\gamma_1+{\rm i}\beta_1)+{\rm i}\sin(D+\beta_1-{\rm i}\gamma_1)}{\cosh(C+\gamma_1)+
    \frac{\sin(\beta_1)}{\cosh(\gamma_1)}\sin(D+\beta_1)}\right]e^{{\rm i}\theta_1},
\end{equation*}
$q_2[1]=\frac{c_2}{c_1}q_1[1],$ where
\begin{equation*}
    \begin{split}
       C&=\alpha(1-|\zeta_1|^{-2})\zeta_{1i}\left[x+\left(\frac{\alpha}{2}\zeta_{1r}\left(
       (1-|\zeta_1|^{-2})+\frac{(1+|\zeta_1|^{-2})^2}{1-|\zeta_1|^{-2}}\right)-a_1\right)t+e_1\right],  \\
       D&=-\alpha(1+|\zeta_1|^{-2})\zeta_{1r}\left[x+\left(\frac{\alpha}{2}(1-|\zeta_1|^{-2})
       \frac{(\zeta_{1r}^2-\zeta_{1i}^2)}{\zeta_{1r}}-a_1\right)t+f_1\right].\\
    \end{split}
\end{equation*}
We can readily see that (suppose $\zeta_{1i}(1-|\zeta_1|^{-2})>0$)
\begin{equation*}
    \begin{split}
       q_1[1]&\rightarrow c_1e^{2{\rm i}\beta_1+{\rm i}\theta_1},\,\, \text{as }x\rightarrow +\infty,  \\
       q_2[1]&\rightarrow c_2e^{-2{\rm i}\beta_1+{\rm i}\theta_1}, \,\,\text{as }x\rightarrow -\infty.
    \end{split}
\end{equation*}
This kinds of solution can be used to obtain the rogue wave solution through limit technique.
The above breather solution can be reduced to single component NLSE, thus we call it as breather-I to distinguish the above breather-II.

In this subsection, the results can be concluded as follows:
\begin{itemize}
  \item If $a_1=a_2$ and $c_1>c_2$, there exists dark-dark soliton and breather-II soliton. Moreover, if $c_2=0$, the breather solution degenerate as
  bright-dark soliton.
  \item If $a_1=a_2$ and $c_1=c_2$, there is no nontrivial solution.
  \item If $a_1=a_2$ and $c_1<c_2$, there exists breather-I solution and rogue wave solution. But there is no dark-dark soliton.
\end{itemize}

\subsection{Case II: When $a_1\neq a_2$}
In this case, performing the similar way with above subsection, we have the matrix decomposition:
\begin{equation}\label{decom1}
    U_0M=MD_0,\,\, D_0=\mathrm{diag}(\chi-\lambda,\mu-\lambda,\nu-\lambda),
\end{equation}
where $\mathrm{Im}(\chi)\geq\mathrm{Im}(\mu)\geq\mathrm{Im}(\nu),$ $\mathrm{Im}(\chi)\neq\mathrm{Im}(\mu)\neq\mathrm{Im}(\nu),$
\begin{equation*}
M=\begin{bmatrix}
    1 & 1 & 1 \\[8pt]
    \frac{c_1}{\chi+a_1} & \frac{c_1}{\mu+a_1} & \frac{c_1}{\nu+a_1} \\[8pt]
    \frac{c_2}{\chi+a_2} & \frac{c_2}{\mu+a_2} & \frac{c_2}{\nu+a_2} \\
  \end{bmatrix},
\end{equation*}
and $\chi$, $\mu$ and $\nu$ are three different roots of the following cubic equation
\begin{equation}\label{chara}
  \xi^3+(a_1+a_2-2\lambda)\xi^2+[a_1a_2+c_1^2-c_2^2-2(a_1+a_2)\lambda]\xi+a_2c_1^2-a_1c_2^2-2a_1a_2\lambda=0.
\end{equation}
The fundamental solution \eqref{fund} is $\Phi=DMN$, where $N=\mathrm{diag}(e^{{\rm i}A_1},e^{{\rm i}B_1},e^{{\rm i}C_1}),$
\begin{equation*}
    \begin{split}
      A_1&=(\chi-\lambda)x+\left[\frac{1}{2}\chi^2+(c_1^2-c_2^2-\lambda^2)\right]t, \\
      B_1&=(\mu-\lambda)x+\left[\frac{1}{2}\mu^2+(c_1^2-c_2^2-\lambda^2)\right]t,  \\
      C_1&=(\nu-\lambda)x+\left[\frac{1}{2}\nu^2+(c_1^2-c_2^2-\lambda^2)\right]t.
    \end{split}
\end{equation*}
Through the matrix decomposition \eqref{decom1}, we have the following equations:
\begin{equation}\label{imp-eq-1}
\begin{split}
   2\lambda-\chi-\frac{c_1^2}{\chi+a_1}+\frac{c_2^2}{\chi+a_2}=&0 , \\
   2\lambda-\mu-\frac{c_1^2}{\mu+a_1}+\frac{c_2^2}{\mu+a_2}=&0, \\
   2\lambda-\nu-\frac{c_1^2}{\nu+a_1}+\frac{c_2^2}{\nu+a_2}=&0 .
\end{split}
\end{equation}
\begin{thm}
For any $\lambda\in \mathbb{C}_+=\{z|\mathrm{Im}(z)>0\}$, there exists two roots $\chi,\mu\in \mathbb{C}_+$,
 one root $\nu\in\mathbb{C}_-=\{z|\mathrm{Im}(z)<0\}$ for the cubic equation \eqref{chara}.
\end{thm}
\textbf{Proof:}
Since matrix $M^{\dag}JM$ is congruent with matrix $J$. By above equation \eqref{imp-eq-1}, we can obtain
\begin{equation*}
    M^{\dag}JM=\begin{bmatrix}
                                  \frac{2(\bar{\lambda}-\lambda)}{\bar{\chi}-\chi}  & \frac{2(\bar{\lambda}-\lambda)}{\bar{\chi}-\nu}
                                  & \frac{2(\bar{\lambda}-\lambda)}{\bar{\chi}-\nu} \\[8pt]
                    \frac{2(\bar{\lambda}-\lambda)}{\bar{\mu}-\chi}  & \frac{2(\bar{\lambda}-\lambda)}{\bar{\mu}-\mu}&\frac{2(\bar{\lambda}-\lambda)}{\bar{\mu}-\nu}\\
                    \frac{2(\bar{\lambda}-\lambda)}{\bar{\nu}-\chi}&\frac{2(\bar{\lambda}-\lambda)}{\bar{\nu}-\mu} &\frac{2(\bar{\lambda}-\lambda)}{\bar{\nu}-\nu}
                                 \end{bmatrix}
\end{equation*}
which possesses two positive roots and one negative root. We can arrange roots with the order
$\mathrm{Im}(\chi)\geq\mathrm{Im}(\mu)\geq\mathrm{Im}(\nu)$. We merely need to prove that
$\mathrm{Im}(\mu)>0$. It is evident that if $\lambda\neq\bar{\lambda}$, we can not obtain the real root. So $\mathrm{Im}(\mu)\neq0$. We prove it by contradiction. Assuming $\mathrm{Im}(\mu)<0$, we can know that
the matrix
\begin{equation*}
M_1=\begin{bmatrix}
\frac{2(\bar{\lambda}-\lambda)}{\bar{\mu}-\mu}&\frac{2(\bar{\lambda}-\lambda)}{\bar{\mu}-\nu} \\
\frac{2(\bar{\lambda}-\lambda)}{\bar{\nu}-\mu} &\frac{2(\bar{\lambda}-\lambda)}{\bar{\nu}-\nu}
\end{bmatrix}
\end{equation*}
is negative definite. Thus this matrix possesses two negative roots. Together with some simple linear algebra, which implies that
$M^{\dag}JM$ possesses two negative roots and one positive root. A contradiction emerges. Thus we complete the proof.
$\square$

Based on above theorem, one can know that the matrix
\begin{equation*}
    \begin{bmatrix}
  1 & \frac{c_1}{\bar{\chi}+a_1} & \frac{c_2}{\bar{\chi}+a_2} \\
   1 & \frac{c_1}{\bar{\mu}+a_1} & \frac{c_2}{\bar{\mu}+a_2} \\
\end{bmatrix}J\begin{bmatrix}
                     1 & 1 \\
                     \frac{c_1}{\chi+a_1}  &  \frac{c_1}{\mu+a_1}\\
                     \frac{c_2}{\chi+a_2} & \frac{c_2}{\mu+a_2} \\
                   \end{bmatrix}=\begin{bmatrix}
                                  \frac{2(\bar{\lambda}-\lambda)}{\bar{\chi}-\chi}  & \frac{2(\bar{\lambda}-\lambda)}{\bar{\chi}-\mu} \\[8pt]
                    \frac{2(\bar{\lambda}-\lambda)}{\bar{\mu}-\chi}  & \frac{2(\bar{\lambda}-\lambda)}{\bar{\mu}-\mu}
                                 \end{bmatrix}
\end{equation*}
is positive definite. This could be used to construct
the nonsingular solution of mCNLSE \eqref{mvnls}.

We take special solution $\Phi_1$ and $v_1(x,t)$ such that
\begin{equation}\label{ycase3}
    |y_1\rangle=D\begin{bmatrix}
             \varphi_1 \\
             c_1\psi_1 \\
             c_2\phi_1 \\
           \end{bmatrix},\,\,
    \begin{bmatrix}
       \varphi_1 \\
       \psi_1 \\
       \phi_1 \\
    \end{bmatrix}
    =\begin{bmatrix}
                     1 & 1 \\
                     \frac{1}{\chi_1+a_1}  &  \frac{1}{\mu_1+a_1}\\
                     \frac{1}{\chi_1+a_2} & \frac{1}{\mu_1+a_2} \\
                   \end{bmatrix}\begin{bmatrix}
                                  e^{{\rm i}A_1} \\
                                  e^{{\rm i}B_1} \\
                                \end{bmatrix},
\end{equation}
where $A_1=\chi_1[(x+x_1)+\frac{1}{2}\chi_1(t+t_1)]$, $B_1=\mu_1[(x+x_1)+\frac{1}{2}\mu_1(t+t_1)]$. For simplicity, we take $x_1=t_1=0$. Then, we give the following calculations
\begin{equation*}
\begin{split}
   |\varphi_1|^2-c_1^2|\psi_1|^2+c_2^2|\phi_1|^2&=\frac{2(\bar{\lambda}_1-\lambda_1)}{\bar{\chi}_1-\chi_1}e^{{\rm i}(A-\bar{A})}
    Y^{\dag}
  \begin{bmatrix}
                                 1 & \frac{\bar{\chi}_1-\chi_1}{\bar{\chi}_1-\mu_1}  \\
                                 \frac{\bar{\chi}_1-\chi_1}{\bar{\mu}_1-\chi_1} & \frac{\bar{\chi}_1-\chi_1}{\bar{\mu}_1-\mu_1} \\
                               \end{bmatrix}Y, \\
   \bar{\varphi}_1\psi_1&=\frac{e^{{\rm i}(A-\bar{A})}}{\chi_1+a_1}Y^{\dag}\begin{bmatrix}
                                                                                                        1 & \frac{\chi_1+a_1}{\mu_1+a_1} \\
                                                                                                        1 & \frac{\chi_1+a_1}{\mu_1+a_1} \\
                                                                                                      \end{bmatrix}Y, \\
   \bar{\varphi}_1\phi_1&=\frac{e^{{\rm i}(A-\bar{A})}}{\chi_1+a_2}Y^{\dag}\begin{bmatrix}
                                                                                                        1 & \frac{\chi_1+a_2}{\mu_1+a_2} \\
                                                                                                        1 & \frac{\chi_1+a_2}{\mu_1+a_2} \\
                                                                                                      \end{bmatrix}Y,
\end{split}
\end{equation*}
where $Y=\left[1,e^{C_1}\right]^T$, $C_1={\rm i}(B_1-A_1)$. By the formula \eqref{edt}, we have
\begin{equation*}
    \begin{split}
       q_1[1]=&c_1\left(1+\frac{2(\bar{\lambda}_1-\lambda_1)\bar{\varphi}_1\psi_1}{|\varphi_1|^2-c_1^2|\psi_1|^2+c_2^2|\phi_1|^2}\right)e^{{\rm i}\theta_1} , \\
       q_2[1]=&c_2\left(1+\frac{2(\bar{\lambda}_1-\lambda_1)\bar{\varphi}_1\phi_1}{|\varphi_1|^2-c_1^2|\psi_1|^2+c_2^2|\phi_1|^2}\right)e^{{\rm i}\theta_2}.
    \end{split}
\end{equation*}
It follows that the breather solutions of mCNLSE \eqref{mvnls} are
\begin{equation}\label{gene-breather}
\begin{split}
  q_1[1] &= c_1\left(1+\frac{\bar{\chi}_1-\chi_1}{\chi_1+a_1}\frac{1+\frac{\chi_1+a_1}{\mu_1+a_1}e^{2\mathrm{Re}(C_1)}+\frac{\chi_1+a_1}{\mu_1+a_1}e^{C_1}
  +e^{\bar{C}_1}}{1+\frac{\bar{\chi}_1-\chi_1}{\bar{\mu}_1-\mu_1}e^{2\mathrm{Re}(C_1)}+\frac{
  \bar{\chi}_1-\chi_1}{\bar{\chi}_1-\mu_1}e^{C_1}+\frac{\bar{\chi}_1-\chi_1}{\bar{\mu}_1-\chi_1}e^{\bar{C}_1}}\right)e^{{\rm i}\theta_1}, \\
  q_2[2] &= c_2\left(1+\frac{\bar{\chi}_1-\chi_1}{\chi_1+a_2}\frac{1+\frac{\chi_1+a_2}{\mu_1+a_2}e^{2\mathrm{Re}(C_1)}+\frac{\chi_1+a_2}{\mu_1+a_2}
  e^{C_1}+e^{\bar{C}_1}}{1+\frac{\bar{\chi}_1-\chi_1}{\bar{\mu}_1-\mu_1}e^{2\mathrm{Re}(C_1)}+\frac{
  \bar{\chi}_1-\chi_1}{\bar{\chi}_1-\mu_1}e^{C_1}+\frac{\bar{\chi}_1-\chi_1}{\bar{\mu}_1-\chi_1}e^{\bar{C_1}}}\right)e^{{\rm i}\theta_2}.
\end{split}
\end{equation}

\begin{description}
  \item[Rogue wave solution]
It is well known that rogue wave solution can be reduced from certain type breather  solution. In what following,  we give the limit process(when $\chi_1\rightarrow \mu_1$) to obtain the rogue wave solution.
It follows that
\begin{equation*}
\begin{split}
    \frac{|\varphi_1|^2-c_1^2|\psi_1|^2+c_2^2|\phi_1|^2}{2(\bar{\lambda}_1-\lambda_1)}&=
    \frac{e^{{\rm i}(\bar{\chi}_1-\chi_1)[x+\frac{1}{2}(\bar{\chi}_1+\chi_1)t]}}
    {\bar{\chi}_1-\chi_1}Y_1^{\dag}\begin{bmatrix}
                                                    1 & -\frac{1}{1-\frac{\epsilon}{\bar{\chi}_1-\chi_1}} \\[8pt]
                                                   -\frac{1}{1+\frac{\bar{\epsilon}}{\bar{\chi}_1-\chi_1}}  & \frac{1}{1+\frac{\bar{\epsilon}-\epsilon}{\bar{\chi}_1-\chi_1}} \\
                                                  \end{bmatrix}Y_1,
\\
   \bar{\varphi}_1\psi_1&=\frac{e^{{\rm i}(\bar{\chi}_1-\chi_1)[x+\frac{1}{2}(\bar{\chi}_1+\chi_1)t]}}
    {\chi_1+a_1}Y_1^{\dag}\begin{bmatrix}
                                                    1 & -\frac{1}{1+\frac{\epsilon}{\chi_1+a_1}} \\[8pt]
                                                   -1  & \frac{1}{1+\frac{\epsilon}{\chi_1+a_1}}\\
                                                  \end{bmatrix}Y_1,
\\
    \bar{\varphi}_1\phi_1&=\frac{e^{{\rm i}(\bar{\chi}_1-\chi_1)[x+\frac{1}{2}(\bar{\chi}_1+\chi_1)t]}}
    {\chi_1+a_2}Y_1^{\dag}\begin{bmatrix}
                                                    1 & -\frac{1}{1+\frac{\epsilon}{\chi_1+a_2}} \\[8pt]
                                                   -1  & \frac{1}{1+\frac{\epsilon}{\chi_1+a_2}}\\
                                                  \end{bmatrix}Y_1,
\end{split}
\end{equation*}
where $\epsilon=\mu_1-\chi_1$, $Y_1=\begin{bmatrix}
                                                                 1 \\
                                                                 e^{{\rm i}\epsilon(x+
       (\chi_1+\frac{1}{2}\epsilon)t)}  \\
                                                               \end{bmatrix}$.

Taking limit for above equation, we can obtain the rogue wave solution
\begin{equation*}
    \begin{split}
     q_1[1]=&\lim_{\epsilon\rightarrow 0}c_1\left(1+\frac{2(\bar{\lambda}_1-\lambda_1)\bar{\varphi}_1\psi_1}{|\varphi_1|^2-c_1^2|\psi_1|^2+c_2^2|\phi_1|^2}\right)e^{{\rm i}\theta_1},\\
     =&
     c_1\left(1+\frac{-2{\rm i}r_1}{p_1+{\rm i}r_1+a_1}\frac{(x+p_1t)^2+r_1^2t^2+
     \frac{{\rm i}}{p_1+a_1+{\rm i}r_1}(x+p_1t-{\rm i}r_1t)}{(x+p_1t+\frac{1}{2r_1})^2+r_1^2t^2+\frac{1}{4r_1^2}}\right)e^{{\rm i}\theta_1},
\end{split}
\end{equation*}
and
\begin{equation*}
\begin{split}
     q_2[1]=&\lim_{\epsilon\rightarrow 0}c_2\left(1+\frac{2(\bar{\lambda}_1-\lambda_1)\bar{\varphi}_1\phi_1}{|\varphi_1|^2-c_1^2|\psi_1|^2+c_2^2|\phi_1|^2}\right)e^{{\rm i}\theta_2},\\
     =&
     c_2\left(1+\frac{-2{\rm i}r_1}{p_1+{\rm i}r_1+a_2}\frac{(x+p_1t)^2+r_1^2t^2+
     \frac{{\rm i}}{p_1+a_2+{\rm i}r_1}(x+p_1t-{\rm i}r_1t)}{(x+p_1t+\frac{1}{2r_1})^2+r_1^2t^2+\frac{1}{4r_1^2}}\right)e^{{\rm i}\theta_2},
    \end{split}
\end{equation*}
where $p_1=\mathrm{Re}(\chi_1)$, $r_1=\mathrm{Im}(\chi_1)$ and $\chi_1$ is two multiple root.

Furthermore, we can classify the rogue wave solution with four different types by the dynamics behavior.
Since $|q_2[1]|^2$ possesses the similar characteristic with $|q_1[1]|^2$, we merely consider $|q_1[1]|^2$.
We first solve the following equation
\begin{equation*}
    (|q_1[1]|^2)_x=0,\,\,(|q_1[1]|^2)_t=0.
\end{equation*}
Then we have the stationary point
\begin{equation*}
    \begin{split}
      (x,t)&=\left(-\frac{1}{2r_1}, 0\right), \\
      (x,t)&=\left(-\frac{A+(2p_1+a_1)B_1}{2Ar_1},\frac{B_1}{2Ar_1}\right),  \\
      (x,t)&=\left(-\frac{Ar_1+(a_1p_1+p_1^2-r_1^2)B_2}{2Ar_1^2}, \frac{(p_1+a_1)B_2}{2Ar_1}\right),
    \end{split}
\end{equation*}
where
\begin{equation*}
    \begin{split}
       A=&(p_1+a_1)^2+r_1^2,  \\
       B_1=&\pm\sqrt{3(p_1+a_1)^2-r_1^2}, \\
       B_2=&\pm\sqrt{3r_1^2-(p_1+a_1)^2}.
    \end{split}
\end{equation*}
So there are five extreme points when $\frac{1}{3}r_1^2<(p_1+a_1)^2<3r_1^2$, or there are three extreme points.
Another standard for classification of rogue wave solution is the value of
\begin{equation*}
    K=|q_1[1]|^2|_{x=-\frac{1}{2q_1},t=0}=\left[1-\frac{4r_1^2}{(p_1+a_1)^2+r_1^2}\right]^2.
\end{equation*}
When $K>1$, the point is higher than the background; or the point is lower than the background.
Thus we can classify the rogue wave to the following four different types:
\begin{itemize}
  \item If $\frac{(p_1+a_1)^2}{r_1^2}\geq 3$, then the rogue wave is called dark rogue wave.
  \item If $1\leq \frac{(p_1+a_1)^2}{r_1^2}<3$, then the rogue wave is four petals type \cite{zhao}.
  \item If $\frac{1}{3}< \frac{(p_1+a_1)^2}{r_1^2}<1$, then the rogue wave is called two peaks type.
  \item If $\frac{(p_1+a_1)^2}{r_1^2}\leq\frac{1}{3}$, then the rogue wave is called bright rogue wave.
\end{itemize}
It is pointed that similar properties hold for the $|q_2[1]|^2$.

In order to looking for the different types of rogue wave, we give the following methods. For convenience, we set $a_2=-a_1,$ since the system possesses
the Galieo transformation. Suppose $\chi_1=\gamma_1 r_1-a_1+{\rm i}r_1$, where $\gamma_1$ is a real parameter (we can obtain the different type rogue wave through parameter $\gamma_1$), substituting $\chi_1$ into
characteristic equation \eqref{chara}, we have
\begin{equation*}
    r_1^3+br_1^2+cr_1+d=0,
\end{equation*}
where
\begin{equation*}
    \begin{split}
      b=&-\frac{2\lambda+3a_1}{\gamma_1+{\rm i}},  \\
      c=&\frac{4\lambda_1a_1+2a_1^2+c_1^2-c_2^2}{(\gamma_1+{\rm i})^2}, \\
      d=&\frac{-2a_1c_1^2}{(\gamma_1+{\rm i})^3}.
    \end{split}
\end{equation*}
The discriminant for above cubic equation is
\begin{equation*}
    \Delta=18bcd-4b^3d+b^2c^2-4c^3-27d^2=0.
\end{equation*}
Solving above two equations about $c_1^2$ and $c_2^2$, we have
\begin{equation*}
\begin{split}
   c_1^2&=-\frac{(2{\rm i}r_1 -2\lambda_1+2r_1\gamma_1-3a_1)(\gamma_1+{\rm i})^2r_1^2}{2a_1}, \\[8pt]
        c_2^2&=-\frac{(2{\rm i}r_1-2\lambda_1 +2r_1\gamma_1-a_1)(r_1\gamma_1+{\rm i}r_1-2a_1)^2}{2a_1}.
\end{split}
\end{equation*}
Assuming that
$\lambda_1=\lambda_{1r}+{\rm i}\lambda_{1i}$, and solving the equations $\mathrm{Im}(c_1^2)=0$ and $\mathrm{Im}(c_2^2)=0$, we have
\begin{equation*}
    \begin{split}
      \lambda_{1r}=&\frac{(r_1\gamma_1-a_1)(3r_1\gamma_1^2-6a_1\gamma_1+r_1)}{2(r_1\gamma_1^2+r_1-2a_1\gamma_1)},  \\
      \lambda_{1i}=&\frac{r_1^2}{r_1\gamma_1^2+r_1-2a_1\gamma_1}.
    \end{split}
\end{equation*}
Finally, we have
\begin{equation}\label{c1c2}
    \begin{split}
       c_1^2=&\frac{r_1^3(1+\gamma_1^2)^2(r_1\gamma_1-2a_1)}{2a_1[r_1(1+\gamma_1^2)-2a_1\gamma_1]}>0,  \\
       c_2^2=&\frac{\gamma_1[r_1^2+(r_1\gamma_1-2a_1)^2]^2}{2a_1[r_1(1+\gamma_1^2)-2a_1\gamma_1]}>0.
    \end{split}
\end{equation}
Thus the above parameters must satisfy the following condition if $\frac{\gamma_1}{r_1}>0$, then
\begin{equation}\label{condition1}
    \frac{2a_1}{r_1^2}<\frac{\gamma_1}{r_1}<\frac{1+\gamma_1^2}{2a_1},
\end{equation}
or if  $\frac{\gamma_1}{r_1}<0$, then
\begin{equation}\label{condition2}
    \frac{1+\gamma_1^2}{2a_1}<\frac{\gamma_1}{r_1}<\frac{2a_1}{r_1^2}.
\end{equation}

For instance, we give some exact rogue wave solutions without solving the cubic equation \eqref{chara}. Choosing parameters $\gamma_1=1$, $r_1=8$ and $a_1=1$
such that condition \eqref{condition1} or condition \eqref{condition2}, it follows that $p_1=\gamma_1 r_1+a_1=7$ and $|q_1[1]|^2$ is
four petal type, then we can obtain
\begin{equation*}
    \begin{split}
       c_1=&\frac{32}{7}\sqrt{21},  \\
       c_2=&\frac{50}{7}\sqrt{7}.
    \end{split}
\end{equation*}
On the other hand, we have $\frac{1}{3}<\frac{(p_1-a_1)^2}{r_1^2}=\frac{9}{16}<1$. It follows that $|q_2[1]|^2$ is
two-peaks type.
We can plot the above rogue wave solution by soft Maple (Fig. \ref{fig2}).
\begin{figure}[htb]
\centering
\subfigure[$|q_1|^2$]{\includegraphics[height=50mm,width=80mm]{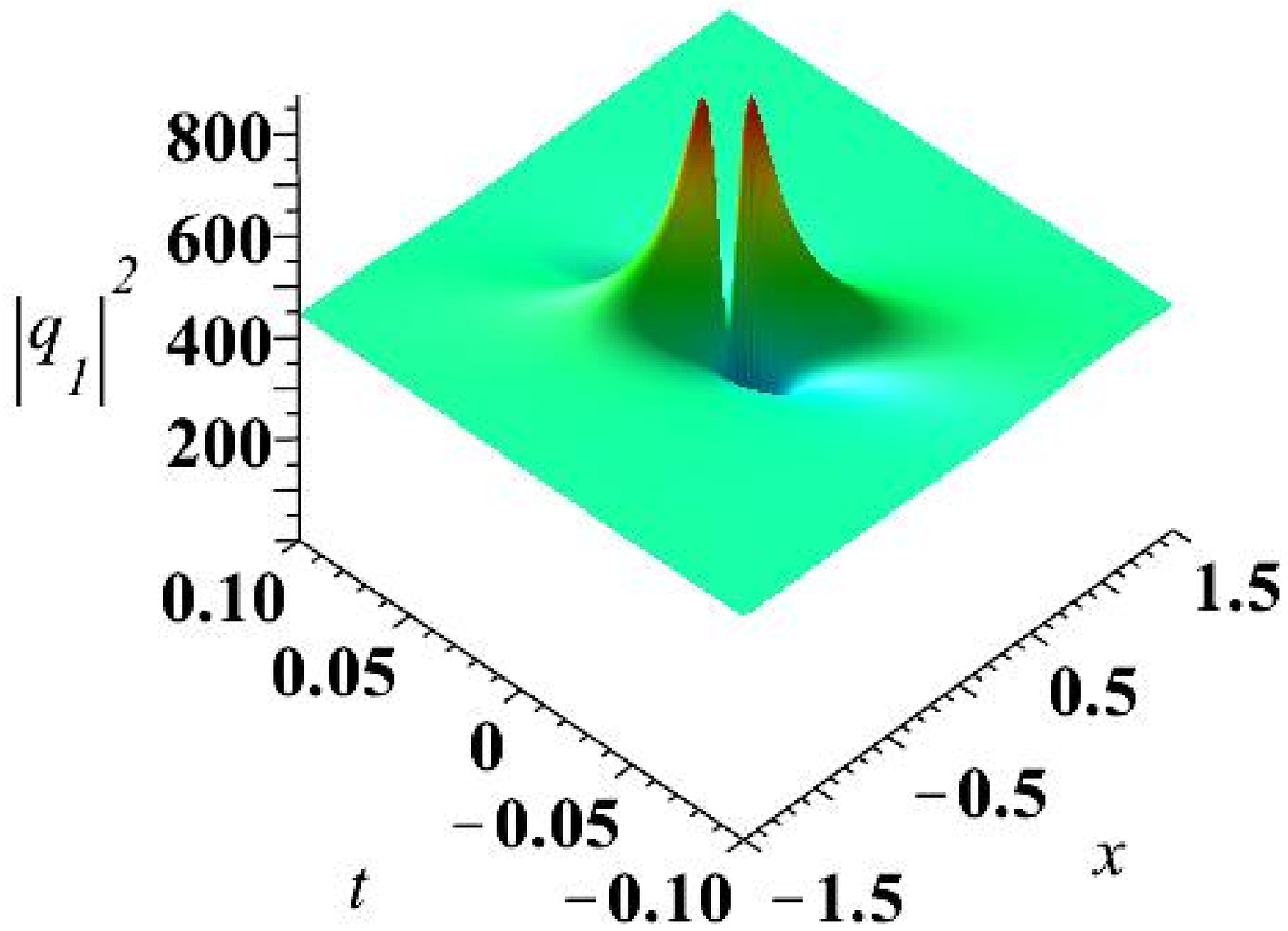}}
\hfil
\subfigure[$|q_2|^2$]{\includegraphics[height=50mm,width=80mm]{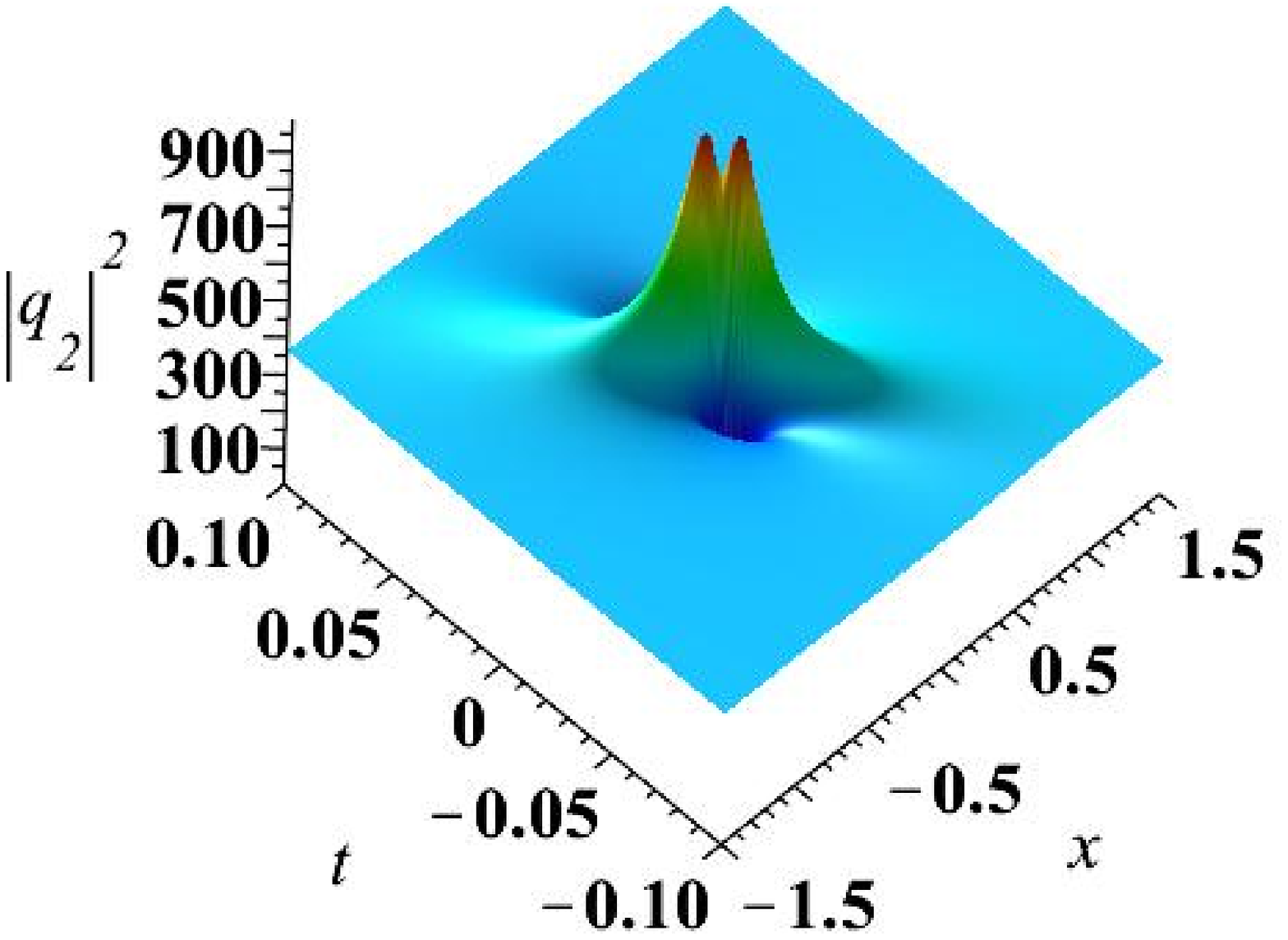}}
\caption{(color online): Rogue wave solution: Parameters $a_1=-a_2=1$, $c_1=\frac{32}{7}\sqrt{21}$, $c_2=\frac{50}{7}\sqrt{7}$, $\lambda_1=\frac{13}{2}+\frac{32}{7}{\rm i}$, $\chi_1=7+8{\rm i}$. It is seen that the solution $|q_1|^2$ possesses four-petals structure and
the solution $|q_2|^2$ possesses two-peaks structure.}\label{fig2}
\end{figure}

Finally, we need to consider the method to determine the multiple roots. We set
\begin{equation*}
    \begin{split}
      a_1&=\alpha+\beta,  \\
      a_2&=\alpha-\beta,
    \end{split}
\end{equation*}
where
\begin{equation}\label{alphabeta}
    \begin{split}
       \alpha&=\frac{a_1+a_2}{2},  \\
       \beta&=\frac{a_1-a_2}{2}.
    \end{split}
\end{equation}
It follows that the matrix $U_0-\xi$ can be represented as
\begin{equation*}
    \begin{bmatrix}
      2\kappa-\zeta & -c_1 & c_2 \\
      c_1 & -\zeta-\beta & 0 \\
      c_2 & 0 & -\zeta+\beta \\
    \end{bmatrix}
\end{equation*}
where $\zeta=\xi+\alpha$, $\kappa=\lambda_1+\frac{1}{2}\alpha$. Then the determinant
of above matrix is
\begin{equation}\label{cubic}
    F=\zeta^3-2\kappa\zeta^2+[c_1^2-c_2^2-\beta^2]\zeta+2\beta^2\kappa-\beta(c_1^2+c_2^2)=0.
\end{equation}
The discriminant of above matrix is
\begin{equation}\label{D(F)}
    D(F)=64\beta^2\left(\kappa^4-\frac{c_1^2+c_2^2}{2\beta}\kappa^3+A_2\kappa^2+A_1\kappa+A_0\right)
\end{equation}
where
\begin{equation*}
    \begin{split}
    A_2&=\left(-\frac{1}{2}\beta^2+\frac{5}{4}(c_2^2-c_1^2)+\frac{1}{16}(c_1^2
    -c_2^2)^2\beta^{-2}\right),   \\
       A_1&=\frac{9}{16}(c_1^2+c_2^2)(2\beta+(c_1^2-c_2^2)\beta^{-1}),  \\
       A_0&=\frac{1}{16}(\beta^2+c_2^2-c_1^2)^3\beta^{-2}
    -\frac{27}{64}(c_1^2+c_2^2)^2.
    \end{split}
\end{equation*}
\begin{thm}\label{thm3}
The quartic equation $D(F)=0$ never possesses two pairs of conjugate complex roots.
\end{thm}
\textbf{Proof:}
We need to analysis the solution of $D(F)=0$. The discriminant of $D(F)=0$ is
\begin{equation*}
    E=-\frac{c_1c_2}{2^{14}\beta^{10}}\left[(4\beta^2+c_2^2-c_1^2)^3+27c_1^2c_2^2(4\beta^2)\right].
\end{equation*}
If $E<0$, then $D(F)=0$ possesses two real roots and a pair of complex conjugate root. If $E>0$, then
then $D(F)=0$ possesses four real roots or two pairs of complex conjugate root. Since $E>0$, then we can deduce that $c_1>c_2$ and the solution of $E=0$ is $\beta^2=\frac{1}{4}(c_1^{2/3}-c_2^{2/3})$.
In what following, we illustrate the equation $D(F)=0$ never possesses two pairs of complex conjugate roots.
If quartic equation has two pairs of complex conjugate roots, then we have
\begin{equation*}
    \begin{split}
      G(\beta^2)= &16A_0\beta^2=(\beta^2+c_2^2-c_1^2)^3
    -\frac{27}{4}\beta^{2}(c_1^2+c_2^2)^2 \\
       =&\beta^6+3(c_2^2-c_1^2)\beta^4-3\left(\frac{5}{2}c_2^2+\frac{1}{2}c_1^2\right)\left(\frac{5}{2}c_1^2+\frac{1}{2}c_2^2\right)\beta^2+(c_2^2-c_1^2)^3>0.
    \end{split}
\end{equation*}
On the other hand, we have
the discriminant of $G(\beta^2)=0$ is
\begin{equation*}
    \Delta=\frac{3^9}{4}c_1^2c_2^2(c_2^2+c_1^2)^4>0,
\end{equation*}
then the equation $G(\beta^2)=0$ possesses three different real roots $\beta_1^2$, $\beta_2^2$ and $\beta_3^2$. By the Vieta formula, we have
\begin{equation*}
    \begin{split}
     \beta_1^2+\beta_2^2+\beta_3^2=&3(c_1^2-c_2^2)>0,  \\
     \beta_1^2\beta_2^2+\beta_2^2\beta_3^2+\beta_3^2\beta_1^2=&-3\left(\frac{5}{2}c_2^2+\frac{1}{2}c_1^2\right)\left(\frac{5}{2}c_1^2+\frac{1}{2}c_2^2\right)<0,  \\
      \beta_1^2\beta_2^2\beta_3^2 =&(c_1^2-c_2^2)^3>0,
    \end{split}
\end{equation*}
it follows that the above equation possesses one positive root and two negative roots. Since $\beta_i^2>0$, then there is merely one positive roots.
The following inequality
\begin{equation*}
    G\left(\frac{1}{4}(c_1^{2/3}-c_2^{2/3})^3\right)=-\frac{27}{16}(c_1^{2/3}-c_2^{2/3})^3\left(\frac{1}{4}(c_1^{2/3}+c_2^{2/3})^6+(c_1^2+c_2^2)^2\right)<0,
\end{equation*}
illustrate that when
$E>0$, then $G(\beta^2)<0$. It follows that when $E>0$, there is four real roots for the equation \eqref{D(F)}. $\square$

Indeed, the quartic equation $D(F)=0$ is condition of multiple roots for the cubic equation \eqref{cubic}. To obtain the rogue wave solution,
another condition is the spectral parameters must be nonreal.
So we can obtain the existence condition of rogue wave is $E<0$ i.e. $a_1\neq a_2$ and $\beta^2>\frac{1}{4}(c_1^{2/3}-c_2^{2/3})^3$, through above theorem.

For the focusing CNLSE, there is three multiple seed for the characteristic equation \eqref{chara}. However, for mixed case or
the defocusing case, there is no three multiple seed. So there is no type-II rogue wave \cite{Guo}. This fact can be verified by the following
elementary fact.
\begin{thm}
The characteristic equation \eqref{chara} never possesses three multiple root.
\end{thm}
\textbf{Proof:}
If characteristic equation \eqref{chara} possesses three multiple seed, then we have
\begin{equation*}
    \zeta^3-2\kappa\zeta^2+[c_1^2-c_2^2-\beta^2]\zeta+2\beta^2\kappa-\beta(c_1^2+c_2^2)=(\zeta-\frac{2\kappa}{3})^3.
\end{equation*}
It follows that
\begin{equation*}
    \begin{split}
      c_1^2-c_2^2-\beta^2-\frac{4}{3}\kappa^2=&0,  \\
      -c_2^2\beta-c_1^2\beta+2\kappa\beta^2+\frac{8}{27}\kappa^3=&0.
    \end{split}
\end{equation*}
Moreover, we have
\begin{equation*}
    \begin{split}
       c_1^2=&\frac{(\frac{2}{3}\kappa+\beta)^3}{2\beta}>0,\,\, \kappa\neq\bar{\kappa},  \\
       c_2^2=&\frac{(\frac{2}{3}\kappa-\beta)^3}{2\beta}>0.
    \end{split}
\end{equation*}
If $\beta>0$, we have $\frac{2}{3}\kappa\pm\beta\in\omega \mathbb{R}^+ \text{ or } \omega^2 \mathbb{R}^+$. If $\beta<0$,
we have $\frac{2}{3}\kappa\pm\beta\in\omega \mathbb{R}^- \text{ or } \omega^2 \mathbb{R}^-$. Indeed, this is no possible. Thus there is no three
multiple root. $\square$

Since the type-II rogue wave solution is obtained by three multiple roots,
there exists not type-II rogue wave solution for the mCNLSE \eqref{mvnls} by above theorem.

  \item[Homoclinical orbits solution]

The homoclinic orbit solution is a kind of space periodical and time exponential decay solution, or called the Akhmediev breather.
When time tends to $\pm\infty$, it tends to a plane wave solution with different phase. From the solution expression of \eqref{gene-breather},
the homoclinic orbit solution can be obtained through choosing parameters $\mathrm{Im}(\chi_1)=\mathrm{Im}(\tau_1)$.

In the following, we present a way to looking for the homoclinic orbit solution for mCNLSE \eqref{mvnls}. Then, it is necessary to analysis the
following characteristic equation \eqref{cubic}
\begin{equation*}
    \zeta^3-2\kappa\zeta^2+(c_1^2-c_2^2-\beta^2)\zeta+2\kappa\beta^2-(c_1^2+c_2^2)\beta=0,
\end{equation*}
where $\kappa=\lambda_1+\frac{1}{2}\alpha$, $\zeta=\tau_1+\alpha$, $\alpha$ and $\beta$ are given in equations \eqref{alphabeta}.
Suppose another root is $\chi_1=\tau_1+\delta$, $\delta\in \mathbb{R}$, then we have
\begin{equation*}
    3\delta \zeta^2+(3\delta^2-4\delta\kappa)\zeta+\delta^3-2\kappa\delta+\delta(c_1^2-c_2^2-\beta^2)=0.
\end{equation*}
It follows that
\begin{equation*}
    \zeta=-\frac{\delta}{2}+\frac{2\kappa}{3}\pm\frac{1}{6}\sqrt{16\kappa^2-3\delta^2+12\beta^2+12(c_2^2-c_1^2)},
\end{equation*}
and it also satisfies the following equation
\begin{equation}\label{chara1}
\begin{split}
 & (16\kappa^2-3\delta^2-12c_1^2+12c_2^2+12\beta^2)\left[3(c_1^2-c_2^2)+3(\delta^2-\beta^2)-4\kappa^2\right]^2 \\
 =& \left[
    16\kappa^3+18(c_2^2-c_1^2-2\beta^2)\kappa+27\beta(c_1^2+c_2^2)\right]^2.
\end{split}
\end{equation}
If we obtain a
pair of conjugate complex roots,  exact values of $\delta$, $\beta$, $c_1$ and $c_2$ are substituted into above equation. Then we can obtain the holoclinic orbit solution through substituting the above mentioned parameters into solutions \eqref{gene-breather}.

Then, we give an explicit example to illustrate the method. Suppose $\alpha=0$, $\beta=1$, $\delta=1$, $c_1=1$ and $c_2=2$, substitute these parameters
into \eqref{chara1} and solve it about $\kappa$, we can obtain that $\kappa=\lambda_1\approx.633263953+1.812212393{\rm i}$. And then substituting above
parameters into characteristic equation \eqref{cubic}, we have
\begin{equation*}
    \begin{split}
      \chi_1= &1.625455953+1.933383832{\rm i},  \\
      \tau_1= &0.625455953+1.933383832{\rm i}.
    \end{split}
\end{equation*}
We give the explicit figure for homoclinic orbit solution by choosing special parameters (Fig. \ref{fig3}).

\begin{figure}[htb]
\centering
\subfigure[$|q_1|^2$]{\includegraphics[height=50mm,width=80mm]{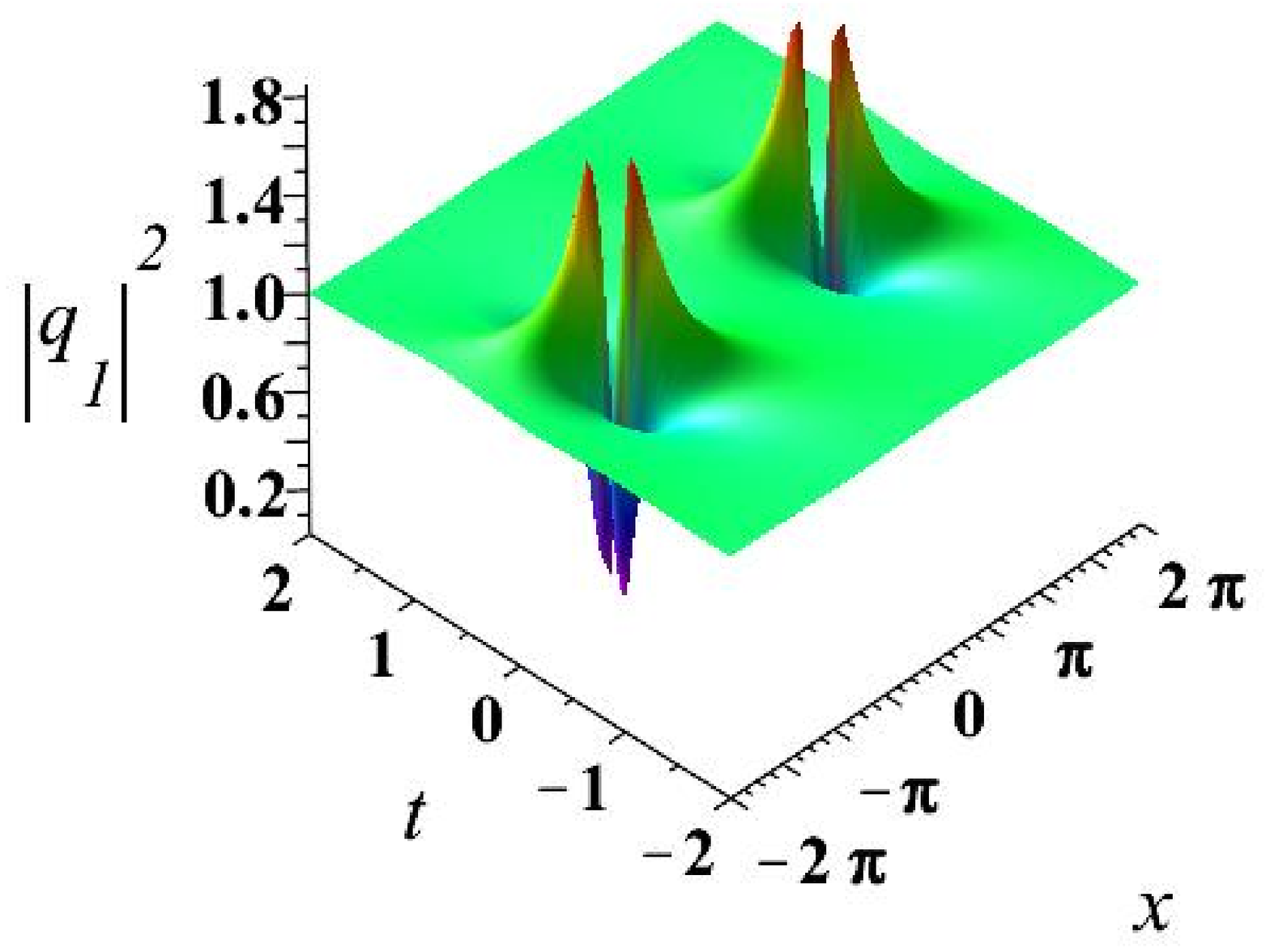}}
\hfil
\subfigure[$|q_2|^2$]{\includegraphics[height=50mm,width=80mm]{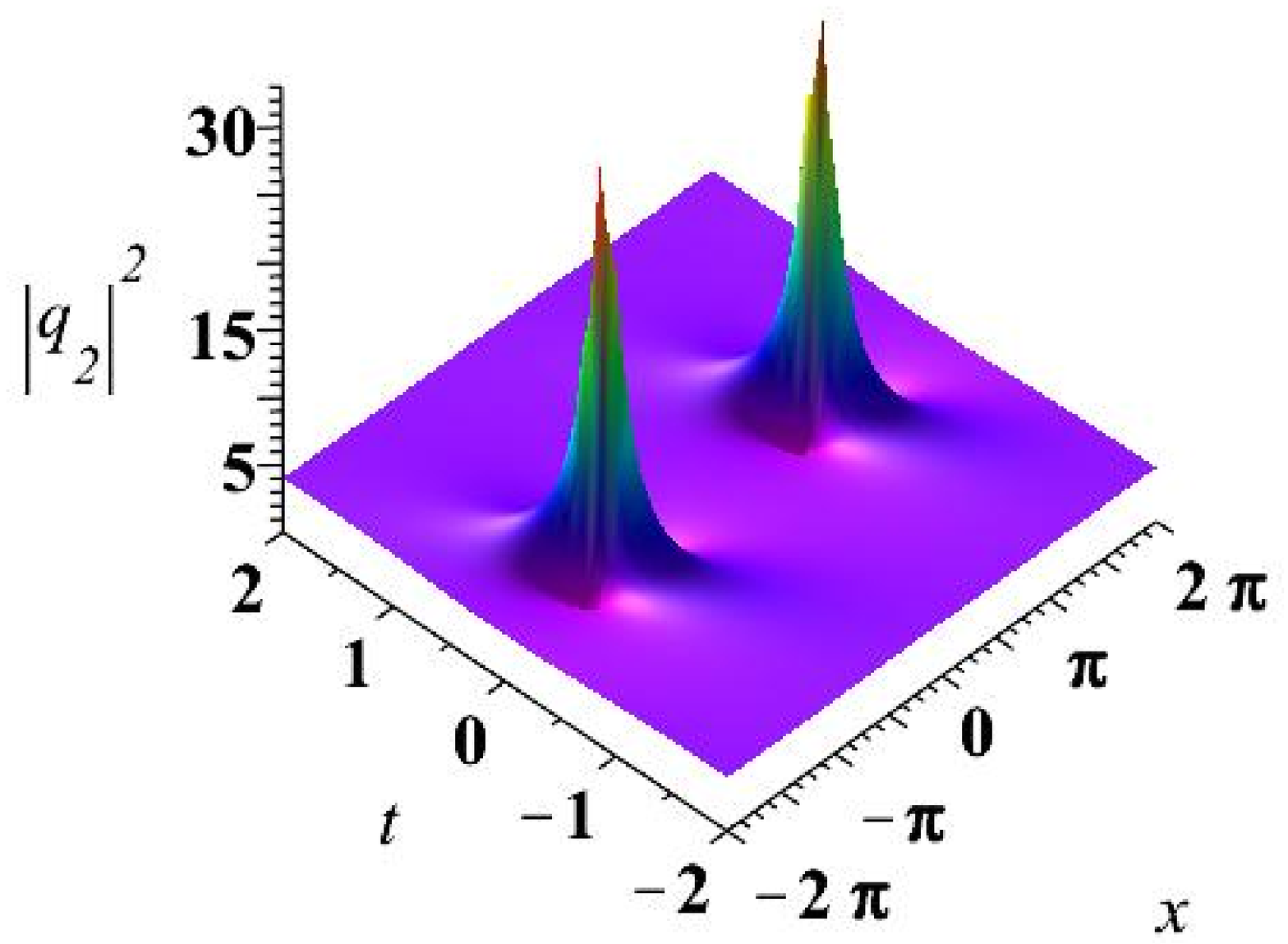}}
\caption{(color online): Homoclinic orbit solution: Parameters $a_1=-a_2=1$, $c_1=1$, $c_2=2$, $\lambda_1=0.633263953+1.812212393{\rm i}$, $\chi_1=
1.625455953+1.933383832{\rm i}$, $\tau_1=.6254559529+1.933383832{\rm i}$. It is seen that the solution $|q_1|^2$ and
$|q_2|^2$ possess the periodical behavior in space.}\label{fig3}
\end{figure}

\end{description}

If $\lambda_1\in \mathbb{R}$, we can obtain the dark-dark soliton solution.  Through \textbf{theorem} \ref{thm3}, the quartic equation
\eqref{D(F)} admits two real roots at least.
So there always exists $\lambda_1$ such that $D(F)<0$, it follows that there is a pair of complex roots for cubic equation \eqref{cubic}. Thus the
dark-dark soliton always exist.
The details for the dark-dark soliton through DT refer to reference \cite{Ling}. Similar as above,
we choose the special solution $\Phi_1$ and $v_1(x,t)$ such that
\begin{equation}\label{ycase4}
    |y_1\rangle=D\begin{bmatrix}
             1&  1  \\[8pt]
             \frac{c_1}{\chi_1+a_1}&\frac{c_1}{\bar{\chi}_1+a_1}  \\[8pt]
             \frac{c_2}{\chi_1+a_2}&\frac{c_2}{\bar{\chi}_1+a_2} \\
           \end{bmatrix} \begin{bmatrix}
           e^{{\rm i}\chi_1(x+\frac{1}{2}\chi_1t)}\\[8pt]
           \alpha_1(\bar{\lambda}_1-\lambda_1)e^{{\rm i}\bar{\chi}_1(x+\frac{1}{2}\bar{\chi}_1t)}  \\
           \end{bmatrix},
\end{equation}
where $\mathrm{Im}(\chi_1)>0.$
Substituting above special solution into \eqref{potential}, we have the dark-dark soliton solution
\begin{equation*}
    \begin{split}
       q_1[1]=&\frac{c_1}{2}\left[1+\frac{\bar{\chi}_1+a_1}{\chi_1+a_1}+\left(1-\frac{\bar{\chi}_1+a_1}{\chi_1+a_1}\right)\tanh(A_1)\right]e^{{\rm i}\theta_1} , \\
       q_2[1]=&\frac{c_2}{2}\left[1+\frac{\bar{\chi}_1+a_2}{\chi_1+a_2}+\left(1-\frac{\bar{\chi}_1+a_2}{\chi_1+a_2}\right)\tanh(A_1)\right]e^{{\rm i}\theta_2},
    \end{split}
\end{equation*}
where $A_1=\mathrm{Im}(\chi_1)[x+\mathrm{Re}(\chi_1)t]+\frac{1}{2}\ln \beta_1$ and $\beta_1=-\mathrm{Im}(\chi_1)\mathrm{Im}\left[\alpha_1\left(1-\frac{c_1^2}{(\bar{\chi}_1+a_1)^2}+\frac{c_2^2}{(\bar{\chi}_1+a_2)^2}\right)\right]>0$.

In summary, the results in this subsection can be concluded as follows:
\begin{itemize}
                 \item If $a_1\neq a_2$ and $\beta^2>\frac{1}{4}(c_1^{2/3}-c_2^{2/3})^3$, there exists rogue wave, breather-I solution and dark-dark
                 soliton solution.
                 \item If $a_1\neq a_2$ and $\beta^2\leq\frac{1}{4}(c_1^{2/3}-c_2^{2/3})^3$, there exists breather-II solution and dark-dark soliton solution.
               \end{itemize}

\section{General localized wave solution formula and interaction between two types of localized wave solution}
The general Darboux matrix for above linear system \eqref{lax} is given in reference \cite{Ling}. We can summarize as the following theorem:
\begin{thm}\label{thm5}
Suppose we have $N$ different vector solutions $\Phi_i$ for linear system \eqref{lax} with $\lambda=\lambda_i$ $(i=1,2,\cdots, N)$, denote $|y_i\rangle=v_i\Phi_i$ ,
where $v_i$ is an appropriate function of $x$ and $t$,
then the $N$-fold DT is
\begin{equation*}
    T_N=I-YM^{-1}(\lambda-S)^{-1}Y^{\dag}J,
\end{equation*}
where
\begin{equation*}
\begin{split}
   Y & =\left[|y_1\rangle,|y_2\rangle,\cdots,|y_N\rangle\right], \\
   S & =\mathrm{diag}\left(\lambda_1,\lambda_2,\cdots,\lambda_N\right),\\
   M & =\left(\frac{\langle y_j|J|y_i\rangle}{\lambda_i-\bar{\lambda}_j}\right)_{N\times N},
\end{split}
\end{equation*}
and $\langle y_i|=|y_i\rangle^{\dag}$, if $\lambda_i=\bar{\lambda}_i$
the element $\frac{\langle y_i|J|y_i\rangle}{\lambda_i-\bar{\lambda}_i}$
is considered as the mean of limit, the transformation between potential functions is
\begin{equation*}
    Q[N]=Q+[\sigma_3,P],\,\, P=YM^{-1}Y^{\dag}J.
\end{equation*}
\end{thm}
In what following, we consider an explicit application of the above theorem.
We choose the seed solution as the plane wave solution \eqref{seed}, then there are three kinds of special solutions for $|y_i\rangle$ (equation \eqref{ycase1}, \eqref{ycase3}
and \eqref{ycase4}), which can be used to construct different types of localized wave solution.
Denote
\begin{equation*}
    |y_i\rangle=\begin{bmatrix}
                  \varphi_i \\
                  c_1e^{{\rm i}\theta_1}\psi_i \\
                  c_2e^{{\rm i}\theta_2}\phi_i \\
                \end{bmatrix}.
\end{equation*}
Through above theorem, we can obtain that the general localized wave solution formula for mCNLS \eqref{mvnls} on the nonzero background:
\begin{equation*}
    \begin{split}
      q_1[N]&=c_1e^{{\rm i}\theta_1}\left(1-2\mathbf{\psi}M^{-1}\mathbf{\varphi}^{\dag}\right), \\
      q_2[N]&=c_2e^{{\rm i}\theta_2}\left(1-2\mathbf{\phi}M^{-1}\mathbf{\varphi}^{\dag}\right),
    \end{split}
\end{equation*}
where
\begin{equation*}
    \begin{split}
     \mathbf{\varphi}=&\left[\varphi_1,\varphi_2,\cdots,\varphi_N\right],  \\
     \mathbf{\psi}=&\left[\psi_1,\psi_2,\cdots,\psi_N\right],  \\
        \mathbf{\phi}=&\left[\phi_1,\phi_2,\cdots,\phi_N\right].
    \end{split}
\end{equation*}
Furthermore, by simple linear algebra formula, setting $\widehat{M}=-\frac{1}{2}M$, we have determinant representation for above general localized wave solution formula:
\begin{equation*}
    \begin{split}
     q_1[N]&=c_1\left(\frac{\det(\widehat{M}+X_1)}{\det(\widehat{M})}\right)e^{i\theta_1},\,\,X_1=\mathbf{\varphi}^{\dag}\mathbf{\psi},  \\
     q_2[N]&=c_2\left(\frac{\det(\widehat{M}+X_2)}{\det(\widehat{M})}\right)e^{i\theta_2},\,\,X_2=\mathbf{\varphi}^{\dag}\mathbf{\phi}.
    \end{split}
\end{equation*}

In what following, we prove that the above general localized wave solutions are nonsingular. We can establish the following theorem:
\begin{thm}
The matrix $\widetilde{M}={\rm i}\widehat{M}$ is negative definite.
\end{thm}
\textbf{Proof:}
We merely give the case when $a_1\neq a_2$, since the case $a_1=a_2$ is similar.
We prove this fact by analysis the elements of matrix $\widetilde{M}$. The elements $|y_i\rangle$ possesses two different choices (here we do not
consider the limit case),
\begin{itemize}
  \item $|y_i\rangle=D\begin{bmatrix}
                      1 & 1 \\[8pt]
                      \frac{c_1}{\chi_i+a_1} &  \frac{c_1}{\mu_i+a_1}  \\[8pt]
                       \frac{c_2}{\chi_i+a_2}  &  \frac{c_2}{\mu_i+a_2}  \\
                    \end{bmatrix}\begin{bmatrix}
                                   e^{{\rm i}\chi_i[x+x_i+\frac{1}{2}\chi_i(t+t_i)]} \\
                                   e^{{\rm i}\mu_i[x+x_i+\frac{1}{2}\mu_i(t+t_i)]} \\
                                 \end{bmatrix}
                    $, $i= 1,2,\cdots, N_1$,\\ $\mathrm{Im}(\chi_i)>0$, $\mathrm{Im}(\mu_i)>0,$ $x_i$, $t_i\in \mathbb{R},$
  \item $|y_j\rangle=D\begin{bmatrix}
                      1 & 1 \\[8pt]
                      \frac{c_1}{\chi_j+a_1} &  \frac{c_1}{\bar{\chi}_j+a_1}  \\[8pt]
                       \frac{c_2}{\chi_j+a_2}  &  \frac{c_2}{\bar{\chi}_j+a_2}  \\
                    \end{bmatrix}\begin{bmatrix}
                                   e^{{\rm i}\chi_j(x+\frac{1}{2}\chi_jt)} \\
                                   \alpha_i(\bar{\lambda}_j-\lambda_j)e^{{\rm i}\bar{\chi}_j(x+\frac{1}{2}\bar{\chi}_jt)} \\
                                 \end{bmatrix}
                    $, $j= N_1+1,N_1+2,\cdots,N_1+N_2$, \\
$N_1+N_2=N$, $\mathrm{Im}(\chi_j)>0$, $\lambda_j\in\mathbb{R}$, $\beta_j=-\mathrm{Im}(\chi_j)\mathrm{Im}\left[\alpha_j\left(1-\frac{c_1^2}{(\bar{\chi}_j+a_1)^2}+\frac{c_2^2}{(\bar{\chi}_j+a_2)^2}\right)\right]>0.$
\end{itemize}
Then we have the following results:
If $i\neq j$, we have (here we merely consider the case $1\leq i\leq N_1$, $N_1+1\leq j\leq N$, the other case can be obtained with a parallel way)
\begin{equation*}
  \begin{split}
    &\frac{\langle y_i|J|y_j\rangle}{2i(\bar{\lambda}_i-\lambda_j)}\\
      =&\begin{bmatrix}
                                   e^{-{\rm i}\bar{\chi}_i\left[x+x_i+\frac{1}{2}\bar{\chi}_i(t+t_i)\right]}, &
                                   e^{-{\rm i}\bar{\mu}_i\left[x+x_i+\frac{1}{2}\bar{\mu}_i(t+t_i)\right]}
                                 \end{bmatrix}\begin{bmatrix}
                     \frac{1}{{\rm i}(\chi_j-\bar{\chi}_i)} \\[8pt]
                      \frac{1}{{\rm i}(\mu_j-\bar{\mu}_i)} \\
                    \end{bmatrix}e^{{\rm i}\chi_j(x+\frac{1}{2}\chi_jt)}\\
      =&-\int^{+\infty}_x \left(e^{-{\rm i}\bar{\chi}_i\left[s+x_i+\frac{1}{2}\bar{\chi}_i(t+t_i)\right]
      +{\rm i}\chi_j(s+\frac{1}{2}\chi_jt)}+e^{-{\rm i}\bar{\mu}_i\left[s+x_i+\frac{1}{2}\bar{\mu}_i(t+t_i)\right]
      +{\rm i}\chi_j(s+\frac{1}{2}\chi_jt)}\right) ds.
  \end{split}
\end{equation*}
Similarly, we have
\begin{equation*}
\begin{split}
  \frac{\langle y_i|J|y_i\rangle}{2i(\bar{\lambda}_i-\lambda_i)}&=-\int_x^{+\infty}\left|
  e^{{\rm i}\chi_i\left[s+x_i+\frac{1}{2}{\chi}_i(t+t_i)\right]}+e^{{\rm i}\mu_i\left[s+x_i+\frac{1}{2}{\mu}_i(t+t_i)\right]}\right|^2  ds,\\
  \frac{\langle y_j|J|y_j\rangle}{2i(\bar{\lambda}_j-\lambda_j)}&=-\int_x^{+\infty}\left|e^{{\rm i}\chi_j(s+\frac{1}{2}\chi_jt)}\right|^2 ds-\frac{\beta_j}{2\mathrm{Im}(\chi_j)}.
\end{split}
\end{equation*}
It follows that, for any nonzero vector $v=(v_1,v_2,\cdots,c_n)^T$, we have
\begin{equation*}
    \begin{split}
      v^{\dag}Mv=&-\int_x^{+\infty}\left|\sum_{i=1}^{N_1}v_i\left(e^{{\rm i}\chi_i\left[s+x_i+\frac{1}{2}{\chi}_i
      (t+t_i)\right]}+e^{{\rm i}\mu_i\left[s+x_i+\frac{1}{2}{\mu}_i(t+t_i)\right]}\right)
      +\sum_{j=1}^{N_2}v_j\left(e^{{\rm i}\chi_j(s+\frac{1}{2}\chi_jt)}\right)\right|^2 ds\\
      &-\left|\sum_{j=1}^{N_2}\frac{v_j\beta_j}{2\mathrm{Im}(\chi_j)}\right|^2<0.
    \end{split}
\end{equation*}
Thus we complete the proof.
$\square$

\subsection{Interaction between two types of localized wave solution}

In this subsection, we consider the cases that different types of localized waves coexist and interplay with each other.
Firstly, we consider the case $a_1\neq a_2$ and
$c_1c_2\neq0$.

Taking
\begin{equation*}
    \begin{bmatrix}
      \varphi_1 \\
      \psi_1 \\
      \phi_1 \\
    \end{bmatrix}=\begin{bmatrix}
                    1&1 \\
                    \frac{1}{\chi_1+a_1}&\frac{1}{\tau_1+a_1} \\
                    \frac{1}{\chi_1+a_2}&\frac{1}{\tau_1+a_2} \\
                  \end{bmatrix}\begin{bmatrix}
                                 e^{{\rm i}\chi_1[(x+x_1)+\frac{1}{2}\chi_1(t+t_1)]} \\
                                 e^{{\rm i}\tau_1[(x+x_1)+\frac{1}{2}\tau_1(t+t_1)]} \\
                               \end{bmatrix}
    ,\end{equation*}
and
\begin{equation}\label{y2}
     \begin{bmatrix}
      \varphi_2 \\
      \psi_2 \\
      \phi_2 \\
    \end{bmatrix}=\begin{bmatrix}
             1&1 \\[8pt]
             \frac{1}{\chi_2+a_1}&\frac{1}{\bar{\chi}_2+a_1} \\[8pt]
             \frac{1}{\chi_2+a_2}&\frac{1}{\bar{\chi}_2+a_2} \\
           \end{bmatrix} \begin{bmatrix}
                                 e^{{\rm i}\chi_2[x+\frac{1}{2}\chi_2t]} \\
                                 \alpha(\bar{\lambda}_2-\lambda_2)e^{{\rm i}\bar{\chi}_2[x+\frac{1}{2}\bar{\chi}_2t]} \\
                               \end{bmatrix},
\end{equation}
where $$\beta=-\mathrm{Im}(\chi_2)\mathrm{Im}\left[\alpha\left(1-\frac{c_1^2}{(\bar{\chi}_2+a_1)^2}+\frac{c_2^2}{(\bar{\chi}_2+a_2)^2}\right)\right]>0,$$ $x_1$ and
$t_1$ are real constants (for simplicity, we set $x_1=t_1=0$),
we can obtain
the general formulas between breather solution and dark-dark soliton solution:
\begin{equation}\label{gene-formula}
    \begin{split}
      q_1[2]=&c_1\left(\frac{\det(M+X_1)}{\det(M)}\right)e^{{\rm i}\theta_1},  \\
      q_2[2]=&c_2\left(\frac{\det(M+X_2)}{\det(M)}\right)e^{{\rm i}\theta_2},
    \end{split}
\end{equation}
where
\begin{equation*}
      M=\begin{bmatrix}
           M_1 & M_2 \\
           M_3 & M_4 \\
         \end{bmatrix},
        \,\,
       X_1=\begin{bmatrix}
           X_{11} & X_{12} \\
           X_{13} & X_{14} \\
         \end{bmatrix},\,\, X_2=\begin{bmatrix}
           X_{21} & X_{22} \\
           X_{23} & X_{24} \\
         \end{bmatrix},
\end{equation*}
and
\begin{equation*}
   \begin{split}
     M_1=&Y_1^{\dag}\begin{bmatrix}
                                                                                              \frac{1}{\bar{\chi}_1-\chi_1}  & \frac{1}{\bar{\chi}_1-\tau_1} \\[8pt]
                                                                                              \frac{1}{\bar{\tau}_1-\chi_1}  & \frac{1}{\bar{\tau}_1-\tau_1} \\
                                                                                             \end{bmatrix}
          Y_1,\,\, Y_1=\begin{bmatrix}
            1 \\ e^{{\rm i}(\tau_1-\chi_1)[x+\frac{1}{2}(\tau_1+\chi_1)t]}\\
          \end{bmatrix},
       \\
     M_2=& Y_1^{\dag}\begin{bmatrix}
                         \frac{1}{\bar{\chi}_1-\chi_2} \\[8pt]
                         \frac{1}{\bar{\tau}_1-\chi_2} \\
                       \end{bmatrix}e^{{\rm i}\chi_2(x+\frac{1}{2}\chi_2t)},
                    \,\, M_3= e^{-{\rm i}\bar{\chi}_2(x+\frac{1}{2}\bar{\chi}_2t)}\begin{bmatrix}
                                                                             \frac{1}{\bar{\chi}_2-\chi_1}, &
                                                                             \frac{1}{\bar{\chi}_2-\tau_1}  \\
                                                                           \end{bmatrix}Y_1, \\
                     M_4=&\beta+\frac{1}{\bar{\chi}_2-\chi_2}e^{-2\mathrm{Im}(\chi_2)(x+\mathrm{Re}(\chi_2)t)}, \\
   \end{split}
\end{equation*}
and
\begin{equation*}
    \begin{split}
       X_{11}=&Y_1^{\dag}\begin{bmatrix}
                                                                                              \frac{1}{\chi_1+a_1}  & \frac{1}{\tau_1+a_1} \\[8pt]
                                                                                               \frac{1}{\chi_1+a_1}  & \frac{1}{\tau_1+a_1}
                                                                                             \end{bmatrix}
          Y_1, \,\,
     X_{12}= Y_1^{\dag}\begin{bmatrix}
                         \frac{1}{\chi_2+a_1} \\[8pt]
                          \frac{1}{\chi_2+a_1}  \\
                       \end{bmatrix}e^{{\rm i}\chi_2(x+\frac{1}{2}\chi_2t)}, \\
      X_{13}=&  e^{-{\rm i}\bar{\chi}_2(x+\frac{1}{2}\bar{\chi}_2t)}\begin{bmatrix}
                                                                             \frac{1}{\chi_1+a_1}, &
                                                                             \frac{1}{\tau_1+a_1}  \\
                                                                           \end{bmatrix}Y_1, \,\,
      X_{14}=\frac{1}{\chi_2+a_1}e^{-2\mathrm{Im}(\chi_2)(x+\mathrm{Re}(\chi_2)t)},
    \end{split}
\end{equation*}
and $X_{2i}=X_{1i}(a_1\rightarrow a_2)$, $i=1,2,3,4.$

\subsubsection{Interaction between dark-dark soliton and breather}
Since $q_2[2]$ is similar with $q_1[2]$, we merely consider the asymptotical behavior for $q_1[2]$.
From above section, we know that $\mathrm{Im}(\chi_1)\geq\mathrm{Im}(\tau_1)>0$ and $\mathrm{Im}(\chi_2)>0$. At the same time, we can know that the
velocity of dark-dark soliton is $v_d=-\mathrm{Re}(\chi_2)$, and the velocity of breather solution is
$$v_b=\frac{\mathrm{Im}(\tau_1)\mathrm{Re}(\tau_1)-\mathrm{Im}(\chi_1)\mathrm{Re}(\chi_1)}{\mathrm{Im}(\chi_1)-\mathrm{Im}(\tau_1)}.$$
We do not consider they possess the same velocity $v_d=v_b$. So we can assume $v_d<v_b$. To analysis the interaction between dark-dark soliton and
breather, we use the standard method of asymptotical analysis.

\begin{figure}[htb]
\centering
\subfigure[$|q_1|^2$]{\includegraphics[height=50mm,width=80mm]{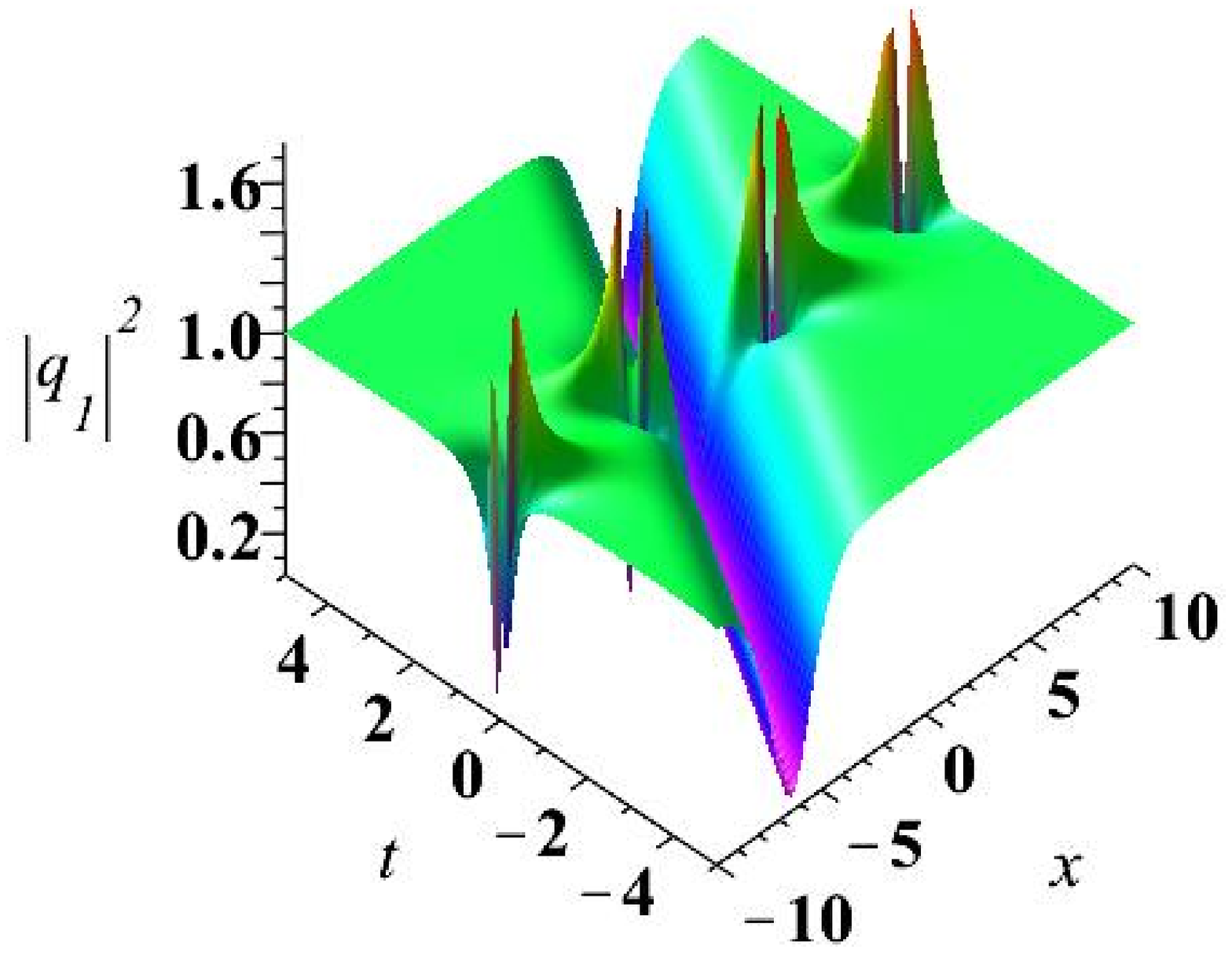}}
\hfil
\subfigure[$|q_2|^2$]{\includegraphics[height=50mm,width=80mm]{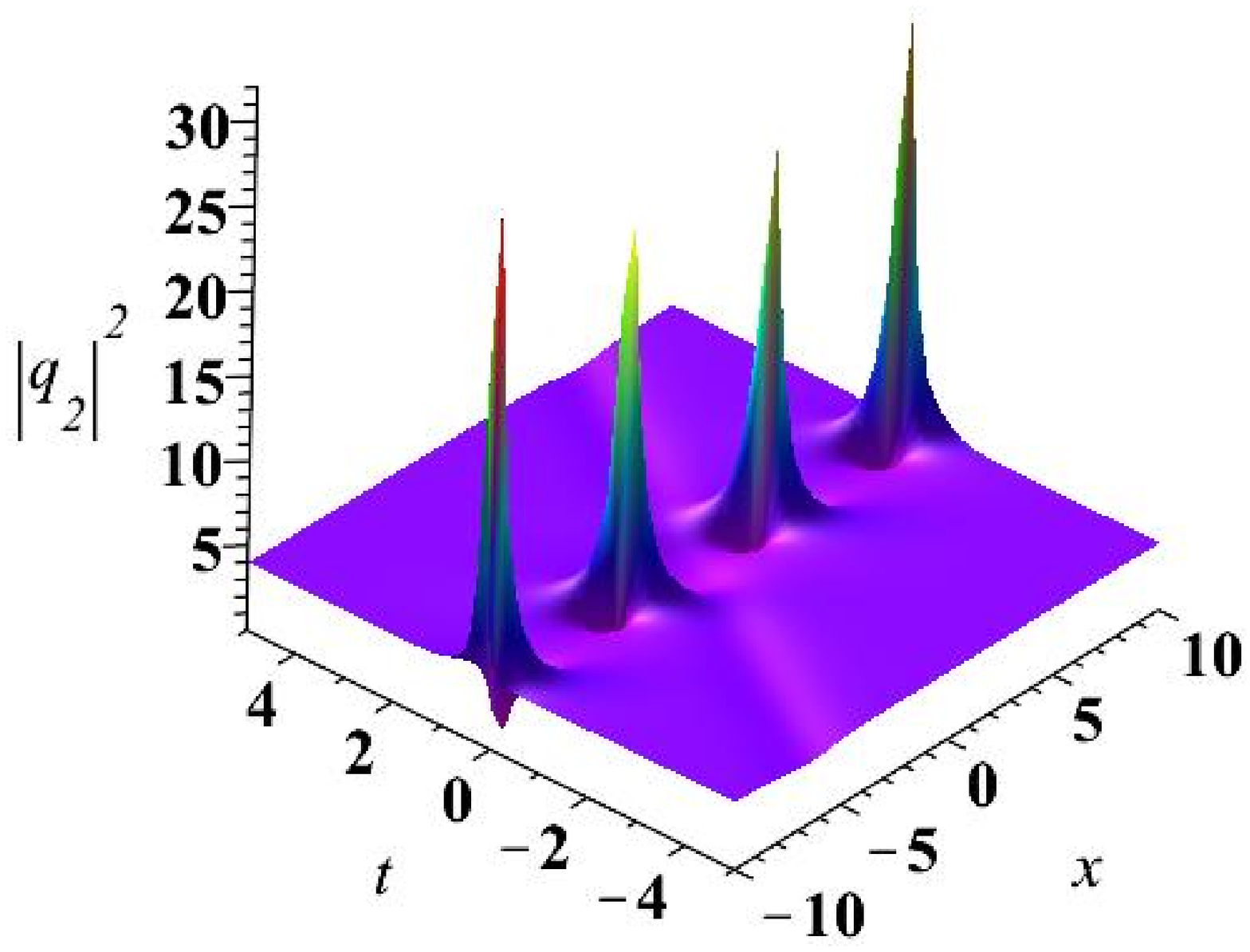}}
\caption{(color online): Dark-dark-breather-I type solution. Parameters $a_1=-a_2=1$, $c_1=1$, $c_2=2$, $\lambda_1=.633263953301716+1.81221239300044{\rm i}$, $\lambda_2
=0$, $\chi_1=1.625455953+1.933383832{\rm i}$, $\tau_1=.6254559529+1.933383832{\rm i}$, $\chi_2=-1.228339172+0.7255696805{\rm i}$, $\beta=0.6891136902{\rm i}$. It is seen that there is homoclinic orbit solution collision with dark-dark soliton, and their interaction is elasticity.}\label{fig4}
\end{figure}

\begin{figure}[htb]
\centering
\subfigure[$|q_1|^2$]{\includegraphics[height=50mm,width=80mm]{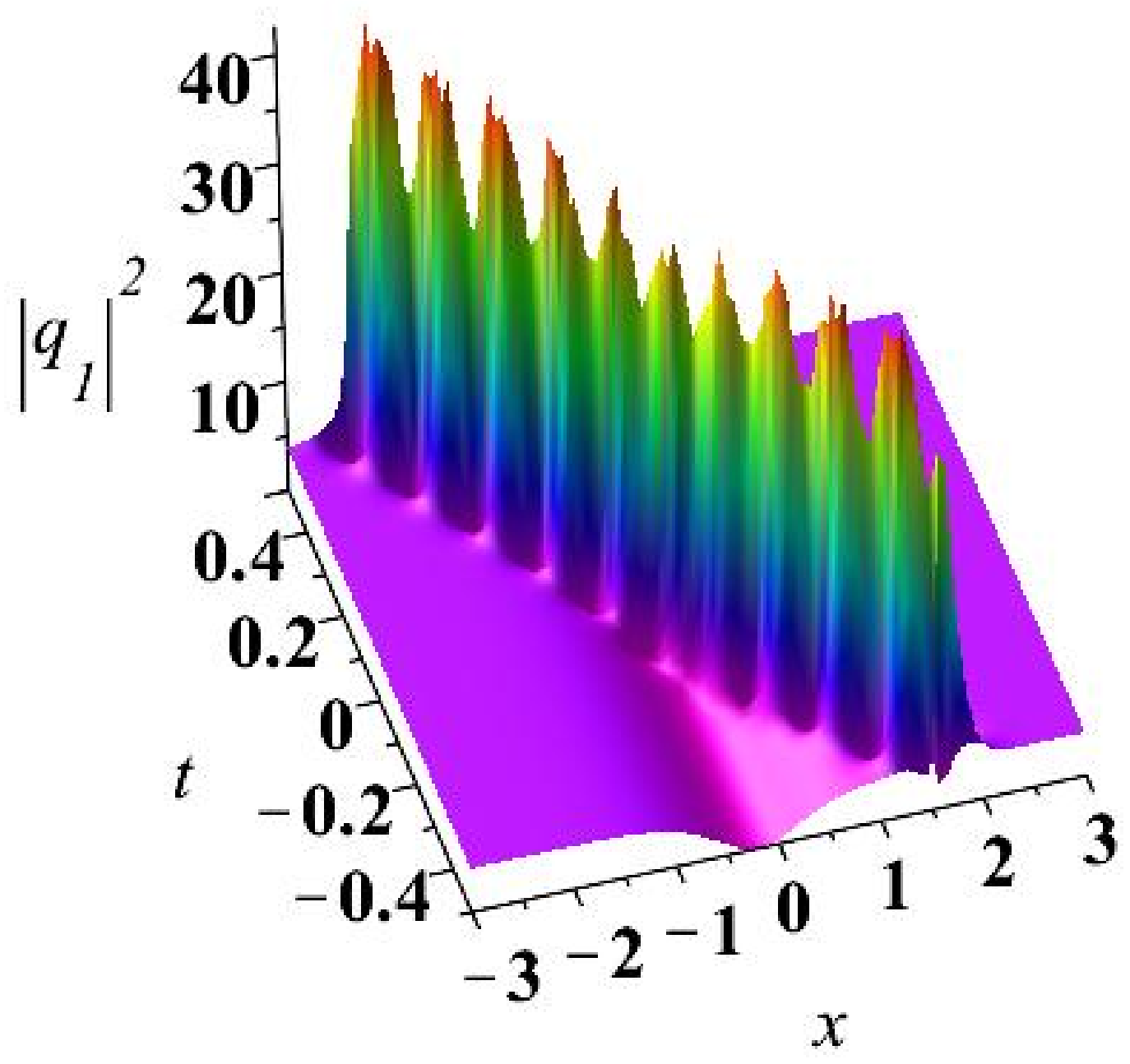}}
\hfil
\subfigure[$|q_2|^2$]{\includegraphics[height=50mm,width=80mm]{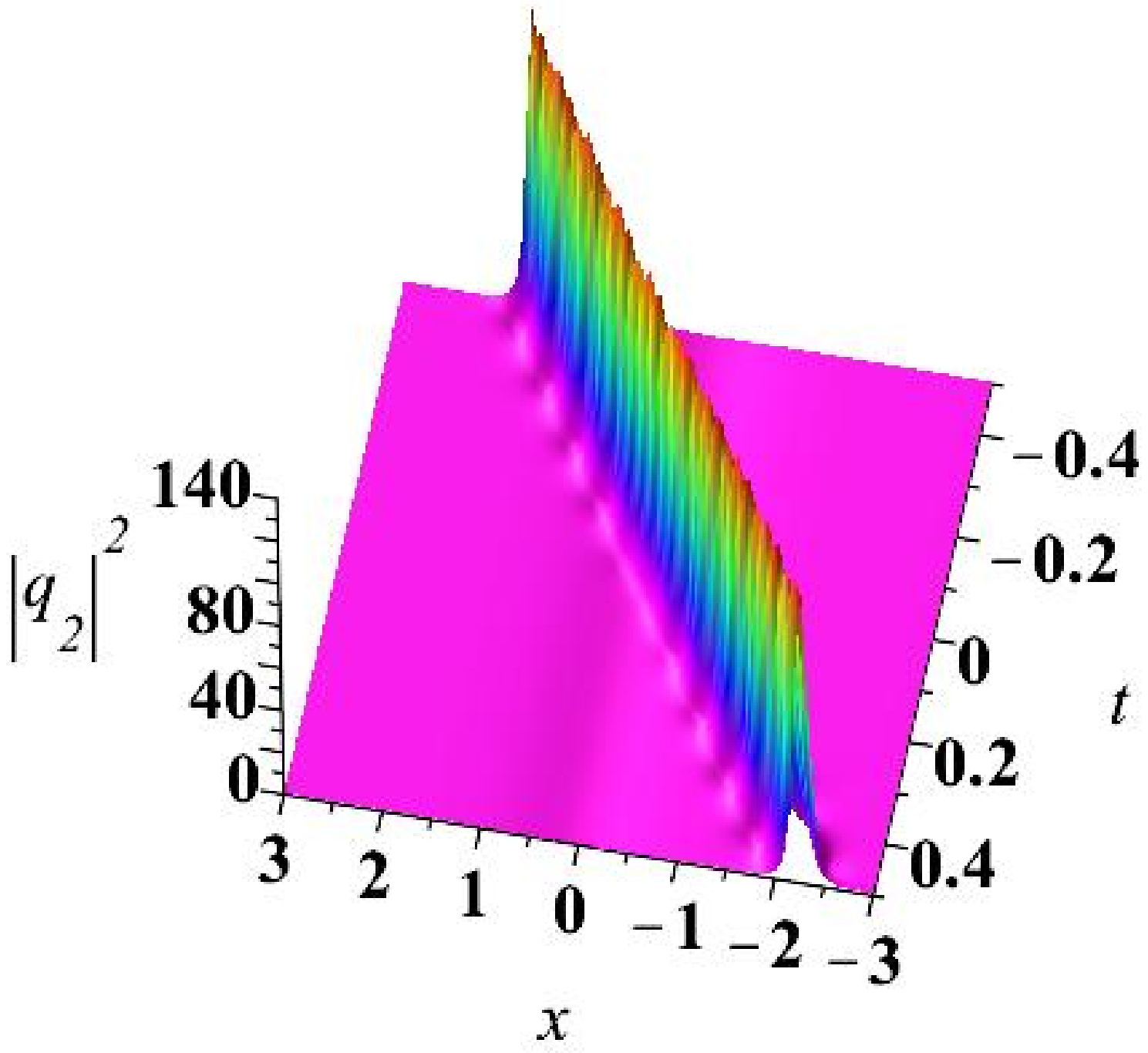}}
\caption{(color online): Dark-dark-breather-II type solution. : Parameters $a_1=-a_2=\frac{1}{20}$, $c_1=2$, $c_2=1$, $\lambda_1=2+5{\rm i}$, $\lambda_2
=0$, $\chi_1=3.901220880+10.25418804{\rm i}$, $\tau_1=0.07993030051+0.01536696563{\rm i}$, $\chi_2=-0.0416053126+1.7328280{\rm i}$, $\beta=0.2885456543{\rm i}$.
It is seen that there is breather-II type solution collision with dark-dark soliton, and their interaction is elasticity too.}\label{fig5}
\end{figure}

Firstly, we fixed $x-v_dt=c_1$, where $c_1$ is a real constant.
\begin{description}
  \item[Case 1] If $t\rightarrow +\infty$, then $\mathrm{Re}({\rm i}(\tau_1-\chi_1)[x+\frac{1}{2}(\tau_1+\chi_1)t])\rightarrow -\infty$, it follows that we have
\begin{equation*}
    q_1[2]\rightarrow c_1 \left(\frac{\bar{\chi}_1+a_1}{\chi_1+a_1}\right)\left[\frac{\frac{\beta}{\bar{\chi}_1-\chi_1}+ \frac{\bar{\chi}_2+a_1}{\chi_2+a_1}A}{\frac{\beta}{\bar{\chi}_1-\chi_1}
    +A}\right]e^{{\rm i}\theta_1},
\end{equation*}
where
\begin{equation*}
    A=\left(\frac{1}{|\bar{\chi}_1-\chi_2|^2}
    +\frac{1}{(\bar{\chi}_1-\chi_1)(\bar{\chi}_2-\chi_2)}\right)e^{-2\mathrm{Im}(\chi_2)(x+\mathrm{Re}(\chi_2)t)}.
\end{equation*}
By the explicit asymptotical expression, we analyze the phase difference between two sides of dark-dark soliton. By simple algebra, we can obtain that
\begin{itemize}
  \item If $\mathrm{Re}({\rm i}\chi_2(x+\frac{1}{2}\chi_2t))
    \rightarrow -\infty,$ then
    $$q_1[2]\rightarrow c_1 e^{{\rm i}(\theta_1+\varphi_1)},\,\,\ln\left(\frac{\bar{\chi}_1+a_1}{\chi_1+a_1}\right)={\rm i}\varphi_1.$$
  \item If $\mathrm{Re}({\rm i}\chi_2(x+\frac{1}{2}\chi_2t))
    \rightarrow +\infty,$ then
    $$q_1[2]\rightarrow c_1 e^{{\rm i}(\theta_1+\varphi_2)},\,\,\ln\left(\frac{\bar{\chi}_1+a_1}{\chi_1+a_1}\frac{\bar{\chi}_2+a_1}{\chi_2+a_1}\right)={\rm i}\varphi_2.$$
\end{itemize}
We can see that the phase between two sides are different.
  \item[Case 2] If $t\rightarrow -\infty$, then $\mathrm{Re}({\rm i}(\tau_1-\chi_1)[x+\frac{1}{2}(\tau_1+\chi_1)t])\rightarrow +\infty$, then we have
\begin{equation*}
    q_1[2]\rightarrow c_1 \left(\frac{\bar{\tau}_1+a_1}{\tau_1+a_1}\right)\left[\frac{\frac{\beta}{\bar{\tau}_1-\tau_1}+ \frac{\bar{\chi}_2+a_1}{\chi_2+a_1}A}
    {\frac{\beta}{\bar{\tau}_1-\tau_1}+A}\right]e^{{\rm i}\theta_1},
\end{equation*}
where
\begin{equation*}
    A=\left(\frac{1}{|\bar{\tau}_1-\chi_2|^2}
    +\frac{1}{(\bar{\tau}_1-\tau_1)(\bar{\chi}_2-\chi_2)}\right)e^{-2\mathrm{Im}(\chi_2)(x+\mathrm{Re}(\chi_2)t)}.
\end{equation*}
Similarly, we can obtain that phase difference between two sides. It follows that
\begin{itemize}
  \item If $\mathrm{Re}({\rm i}\chi_2(x+\frac{1}{2}\chi_2t))
    \rightarrow -\infty,$ then we have
    $$q_1[2]\rightarrow c_1 e^{{\rm i}(\theta_1+\varphi_3)},\,\,\ln\left(\frac{\bar{\tau}_1+a_1}{\tau_1+a_1}\right)={\rm i}\varphi_3.$$
  \item  If $\mathrm{Re}({\rm i}\chi_2(x+\frac{1}{2}\chi_2t))
    \rightarrow +\infty,$ we have
    $$q_1[2]\rightarrow c_1 e^{{\rm i}(\theta_1+\varphi_4)},\,\,
    \ln\left(
    \frac{\bar{\chi}_2+a_1}{\chi_2+a_1}\frac{\bar{\tau}_1+a_1}{\tau_1+a_1}\right)={\rm i}\varphi_4.$$
\end{itemize}
\end{description}
Secondly, we fixed $x-v_bt=c_2$, where $c_2$ is a real constant.
\begin{description}
  \item[Case 3] If $t\rightarrow -\infty$, then $\mathrm{Re}({\rm i}\chi_2(x+\frac{1}{2}\chi_2t))\rightarrow +\infty$, it follows that
  \begin{equation*}
    q_1[2]\rightarrow c_1\left(\frac{\bar{\chi}_2+a_1}{\chi_2+a_1}\right)\left[\frac{\frac{\bar{\chi}_1+a_1}{\chi_1+a_1}B_1
    +\frac{\bar{\chi}_1+a_1}{\tau_1+a_1}B_2+\frac{\bar{\tau}_1+a_1}{\chi_1+a_1}B_3+\frac{\bar{\tau}_1+a_1}{\tau_1+a_1}B_4}{B_1+B_2+B_3+B_4}\right]e^{{\rm i}
    \theta_1},
  \end{equation*}
where
\begin{equation*}
    \begin{split}
       B_1=&\frac{1}{(\bar{\chi}_2-\chi_2)(\bar{\chi}_1-\chi_1)}+\frac{1}{|\bar{\chi}_1-\chi_2|^2},  \\
       B_2=&\left[\frac{1}{(\bar{\chi}_2-\chi_2)(\bar{\chi}_1-\tau_1)}-\frac{1}{(\bar{\chi}_1-\chi_2)
       (\bar{\chi}_2-\tau_1)}\right]e^{-{\rm i}(\bar{\tau}_1-\bar{\chi}_1)[x+\frac{1}{2}(\bar{\tau}_1+\bar{\chi}_1)t]},  \\
      B_3=&\left[\frac{1}{(\bar{\chi}_2-\chi_2)(\bar{\tau}_1-\chi_1)}-\frac{1}{(\bar{\tau}_1-\chi_2)
       (\bar{\chi}_2-\chi_1)}\right]e^{{\rm i}(\tau_1-\chi_1)[x+\frac{1}{2}(\tau_1+\chi_1)t]},   \\
      B_4=&\left[\frac{1}{(\bar{\chi}_2-\chi_2)(\bar{\tau}_1-\tau_1)}+\frac{1}{
       |\bar{\chi}_2-\tau_1|^2}\right]e^{2\mathrm{Re}\left({\rm i}(\tau_1-\chi_1)[x+\frac{1}{2}(\tau_1+\chi_1)t]\right)}.
    \end{split}
\end{equation*}
Similar as case 1, we can obtain that phase difference between two sides. It follows that
\begin{itemize}
  \item If $\mathrm{Re}\left({\rm i}(\tau_1-\chi_1)[x+\frac{1}{2}(\tau_1+\chi_1)t]\right)\rightarrow-\infty$, then
\begin{equation*}
    q_1[2]\rightarrow c_1 e^{{\rm i}(\theta_1+\varphi_2)}.
\end{equation*}
  \item If $\mathrm{Re}\left({\rm i}(\tau_1-\chi_1)[x+\frac{1}{2}(\tau_1+\chi_1)t]\right)\rightarrow+\infty$, then
\begin{equation*}
    q_1[2]\rightarrow c_1 e^{{\rm i}(\theta_1+\varphi_4)}.
\end{equation*}
\end{itemize}
  \item[Case 4] If $t\rightarrow +\infty$, then $\mathrm{Re}\left({\rm i}\chi_2(x+\frac{1}{2}\chi_2t)\right)\rightarrow -\infty$, it follows that
  \begin{equation*}
   q_1[2]\rightarrow c_1\left[\frac{\frac{\bar{\chi}_1+a_1}{\chi_1+a_1}B_1
    +\frac{\bar{\chi}_1+a_1}{\tau_1+a_1}B_2+\frac{\bar{\tau}_1+a_1}{\chi_1+a_1}B_3+\frac{\bar{\tau}_1+a_1}{\tau_1+a_1}B_4}{B_1+B_2+B_3+B_4}\right]e^{{\rm i}
    \theta_1},
  \end{equation*}
where
\begin{equation*}
    \begin{split}
      B_1=&\frac{1}{\bar{\chi}_1-\chi_1},  \,\,
      B_2=\frac{1}{\bar{\chi}_1-\tau_1}e^{-{\rm i}(\bar{\tau}_1-\bar{\chi}_1)[x+\frac{1}{2}(\bar{\tau}_1+\bar{\chi}_1)t]},\\
      B_3=&\frac{1}{\bar{\tau}_1-\chi_1}e^{{\rm i}(\tau_1-\chi_1)[x+\frac{1}{2}(\tau_1+\chi_1)t]},\\
      B_4=&\frac{1}{\bar{\tau}_1-\tau_1}e^{2\mathrm{Re}\left({\rm i}(\tau_1-\chi_1)[x+\frac{1}{2}(\tau_1+\chi_1)t]\right)}.
    \end{split}
\end{equation*}
Similar as case 1, we can obtain that phase difference between two sides. It follows that
\begin{itemize}
  \item If $\mathrm{Re}\left({\rm i}(\tau_1-\chi_1)[x+\frac{1}{2}(\tau_1+\chi_1)t]\right)\rightarrow-\infty$, then
\begin{equation*}
    q_1[2]\rightarrow c_1 e^{{\rm i}(\theta_1+\varphi_1)}.
\end{equation*}
  \item If $\mathrm{Re}\left({\rm i}(\tau_1-\chi_1)[x+\frac{1}{2}(\tau_1+\chi_1)t]\right)\rightarrow+\infty$, then
\begin{equation*}
    q_1[2]\rightarrow c_1 e^{{\rm i}(\theta_1+\varphi_3)}.
\end{equation*}
\end{itemize}
\end{description}
Finally, we can see that the asymptotical analysis between case 1,case 2 and case 3, case 4 is consistence.
To demonstrate the dynamics of this type solution explicitly, we plot some figures with setting related parameters by Maple. In Fig \ref{fig4}, we show the
interaction between one homoclinic orbit solution and dark-dark solution. In Fig \ref{fig5}, we show the
interaction between one breather-II type solution and dark-dark solution.

\subsubsection{Dark-dark-rogue solution}
\begin{figure}[htb]
\centering
\subfigure[$|q_1|^2$]{\includegraphics[height=50mm,width=80mm]{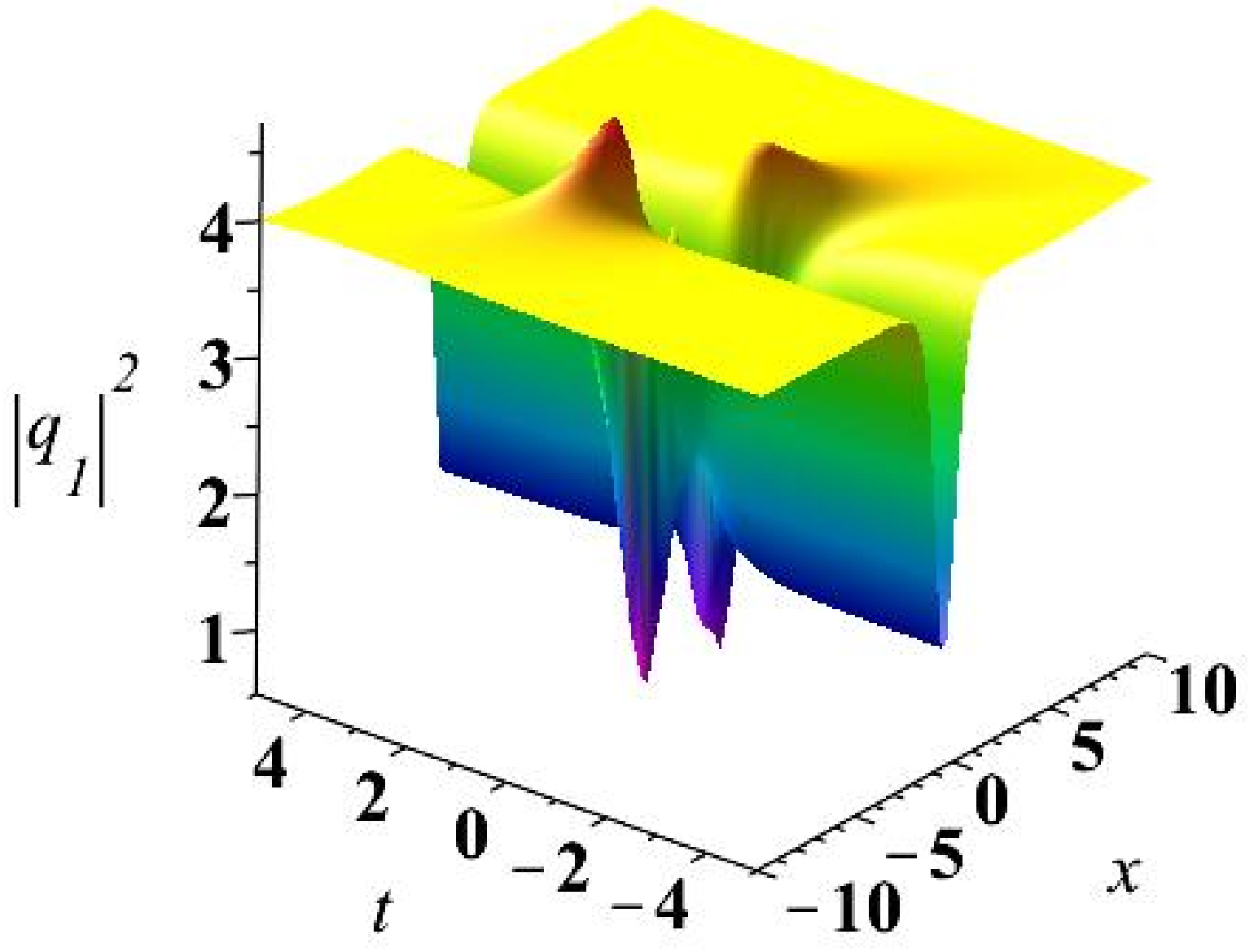}}
\hfil
\subfigure[$|q_2|^2$]{\includegraphics[height=50mm,width=80mm]{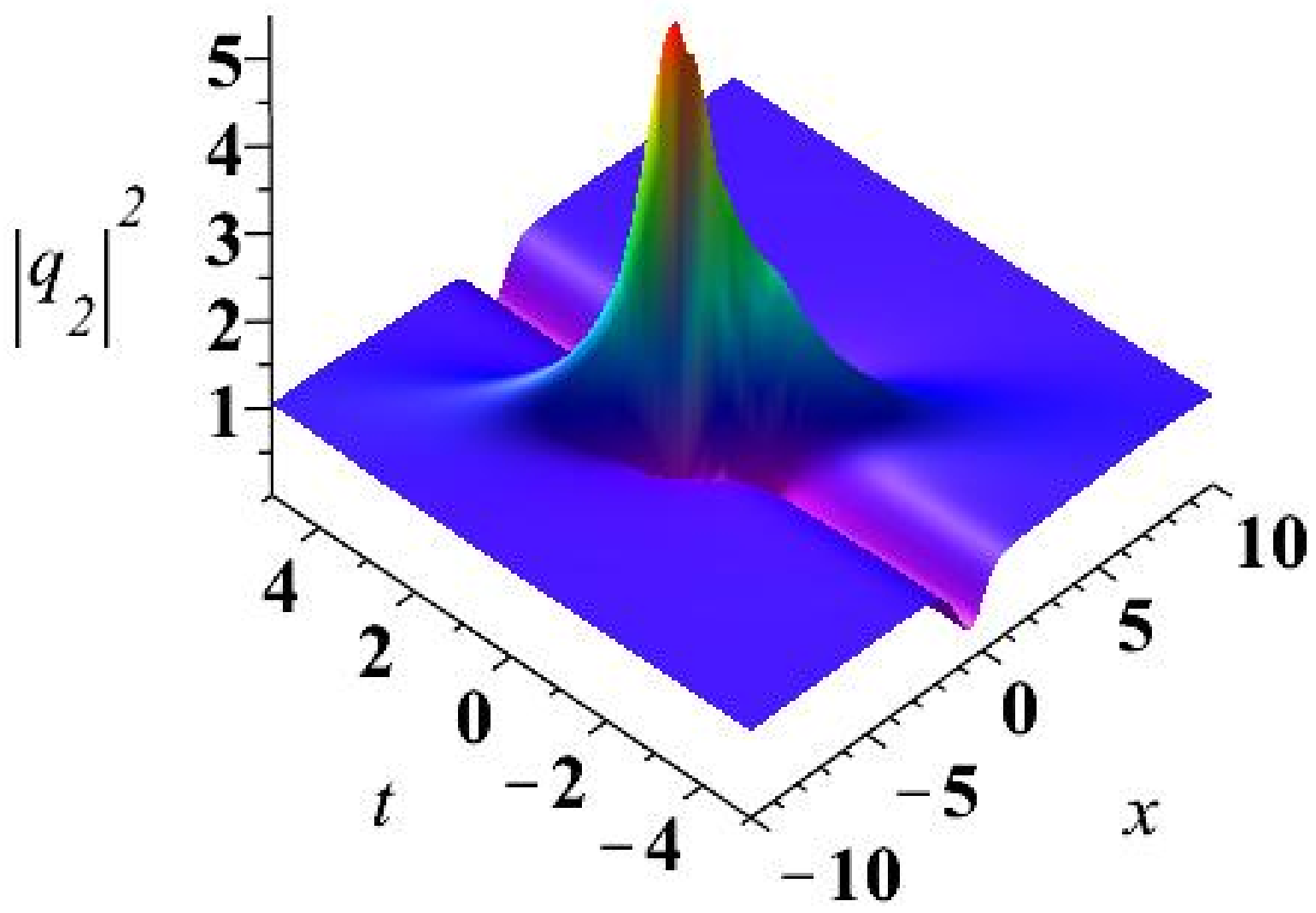}}
\caption{(color online): Dark-dark-rogue solution: Parameters $a_1=-a_2=1$, $c_1=2$, $c_2=1$, $\lambda_1=1.24185466772002+.636002000756738{\rm i}$, $\lambda_2
=0.8333333333$, $\chi_1=1.356709486+1.087820879{\rm i}$, $\chi_2=1.414213562{\rm i}$, $\beta=.3535533907{\rm i}$. It is seen that
 there is dark-dark rogue wave solution collision with rogue wave solution.}\label{fig6}
\end{figure}

In this paragraph, we analysis the interaction between dark-dark soliton and rogue wave solution. To
derive the dark-dark-rogue solution, we use the limit technique $\tau_1\rightarrow \chi_1$.
Denote $\epsilon=\tau_1-\chi_1$, we can have
the following expression
\begin{equation*}
    \begin{split}
      \frac{1}{\bar{\chi}_1-\chi_1-\epsilon}=&\frac{1}{\bar{\chi}_1-\chi_1}+\frac{\epsilon}{(\bar{\chi}_1-\chi_1)^2}+o(\epsilon^2),  \\
      \frac{1}{\bar{\epsilon}+\bar{\chi}_1-\chi_1}=&\frac{1}{\bar{\chi}_1-\chi_1}-\frac{\bar{\epsilon}}{(\bar{\chi}_1-\chi_1)^2}+o(\bar{\epsilon}^2),   \\
      \frac{1}{\bar{\chi}_1-\chi_1-\epsilon+\bar{\epsilon}}=&\frac{1}{\bar{\chi}_1-\chi_1}+\frac{\epsilon-\bar{\epsilon}}{(\bar{\chi}_1-\chi_1)^2}-\frac{2\epsilon
      \bar{\epsilon}}{(\bar{\chi}_1-\chi_1)^3}+o(\epsilon^2,\bar{\epsilon}^2),  \\
      \frac{1}{\bar{\epsilon}+\bar{\chi}_1-\chi_2}=&\frac{1}{\bar{\chi}_1-\chi_2}-\frac{\bar{\epsilon}}{(\bar{\chi}_1-\chi_2)^2}+o(\bar{\epsilon}^2),\\
      \frac{1}{\bar{\chi}_2-\chi_1-\epsilon}=&\frac{1}{\bar{\chi}_2-\tau_1}+\frac{\epsilon}{(\bar{\chi}_2-\tau_1)^2}+o(\epsilon^2),\\
      \frac{1}{\epsilon+\chi_1+a_1}=&\frac{1}{\chi_1+a_1}-\frac{\epsilon}{(\chi_1+a_1)^2}+o(\epsilon^2),
    \end{split}
\end{equation*}
and
\begin{equation*}
    \begin{split}
      e^{{\rm i}\epsilon(x+(\chi_1+\frac{1}{2}\epsilon)t)}=&1+{\rm i}\epsilon(x+\chi_1t)+o(\epsilon^2),  \\
      e^{-{\rm i}\bar{\epsilon}(x+(\bar{\chi}_1+\frac{1}{2}\bar{\epsilon})t)}=&1-{\rm i}\bar{\epsilon}(x+\bar{\chi}_1t)+o(\bar{\epsilon}^2),\\
      e^{{\rm i}\epsilon(x+(\chi_1+\frac{1}{2}\epsilon)t)-{\rm i}\bar{\epsilon}(x+(\bar{\chi}_1+\frac{1}{2}\bar{\epsilon})t)}&=1+{\rm i}\epsilon(x+\chi_1t)-
      {\rm i}\bar{\epsilon}(x+\bar{\chi}_1t)+\frac{\epsilon\bar{\epsilon}}{2}|x+\chi_1t|^2+o(\epsilon^2,\bar{\epsilon}^2).
    \end{split}
\end{equation*}
By the following expansion and \eqref{gene-formula}, we can obtain the dark-dark-rogue solution
\begin{equation}\label{dark-rogue}
    q_1[2]=c_1\left(\frac{F_1+F_2F_3}{D_1+D_2+D_3}\right)e^{{\rm i}\theta_1},\,\, q_2[2]=c_2\left(\frac{G_1+G_2G_3}{D_1+D_2+D_3}\right)e^{{\rm i}\theta_2},
\end{equation}
where
\begin{equation*}
    \begin{split}
       D_1=&\frac{\beta}{\bar{\chi}_1-\chi_1}\left(|x+\chi_1t|^2-\frac{1}{(\bar{\chi}_1-\chi_1)^2}\right),  \\
       D_2=&\frac{1}{(\bar{\chi}_1-\chi_1)(\bar{\chi}_2-\chi_2)}\left(|x+\chi_1t|^2-\frac{1}{(\bar{\chi}_1-\chi_1)^2}\right)
       e^{-2\mathrm{Im}(\chi_2)[x+\mathrm{Re}(\chi_2)t]}, \\
       D_3=&\frac{1}{(\bar{\chi}_1-\chi_2)(\bar{\chi}_2-\chi_1)}\left({\rm i}(x+\bar{\chi}_1t)+\frac{1}{\bar{\chi}_1-\chi_2}\right)\left(
       {\rm i}(x+\chi_1t)+\frac{1}{\bar{\chi}_2-\chi_1}\right),
    \end{split}
\end{equation*}
and
\begin{equation*}
    \begin{split}
      F_1=&\left(\frac{(\bar{\chi}_1+a_1)|x+\chi_1t|^2}{(\chi_1+a_1)(\bar{\chi}_1-\chi_1)}-\frac{1}{(\bar{\chi}_1-\chi_1)^3}\right)
      \left(\beta+\frac{(\bar{\chi}_1+a_1)e^{-2\mathrm{Im}(\chi_2)[x+\mathrm{Re}(\chi_2)t]}}{(\chi_1+a_1)(\bar{\chi}_1-\chi_1)}\right),  \\
      F_2=&\frac{{\rm i}(x+\bar{\chi}_1t)}{\chi_2+a_1}+\left(\frac{1}{(\bar{\chi}_1^2-\chi_2)^2}+
      \frac{{\rm i}(x+\bar{\chi}_1t)}{\bar{\chi}_1-\chi_1}\right)e^{{\rm i}\chi_2(x+\frac{1}{2}\chi_2t)},\\
      F_3=&\frac{{\rm i}(x+\chi_1t)}{\chi_1+a_1}-\frac{1}{(\chi_1+a_1)^2}+\left(\frac{1}{(\bar{\chi}_2-\chi_1)^2}
      +\frac{{\rm i}(x+\chi_1t)}{\bar{\chi}_2-\chi_1}\right)e^{-{\rm i}\bar{\chi}_2(x+\frac{1}{2}\bar{\chi}_2t)},
    \end{split}
\end{equation*}
and $G_i=F_i(a_1\rightarrow a_2)$, $i=1,2,3$. The asymptotical analysis of the above subsubsection is still valid for the dark-dark-rogue solution, since
the dark-dark-rogue solution is nothing but the limit for solution \eqref{gene-formula}. Finally we show the explicit dynamics by plotting figure (Fig. \ref{fig6}).

\subsection{Two dark-one bright soliton solution}
In this subsection, we consider the special case $a_1=a_2$.
Firstly, we consider $c_1>c_2>0$, choosing the following special solution
 $$|y_1\rangle=\begin{bmatrix}
   e^{{\rm i}(\chi_1x+\frac{1}{2}\chi_1^2t)} \\
  \frac{c_1}{\chi_1+a_1} e^{{\rm i}(\chi_1x+\frac{1}{2}\chi_1^2t)}+c_2\alpha_1e^{{\rm i}(-a_1x+\frac{1}{2}a_1^2t)} \\
   \frac{c_2}{\chi_1+a_1}  e^{{\rm i}(\chi_1x+\frac{1}{2}\chi_1^2t)}+c_1\alpha_1e^{{\rm i}(-a_1x+\frac{1}{2}a_1^2t)} \\
  \end{bmatrix}$$ and $|y_2\rangle$ \eqref{y2} as above subsection, for simplicity we take $\alpha_1=1$, we can obtain the solution \eqref{gene-formula}
in above subsection with
\begin{equation*}
   \begin{split}
     M_1=&\frac{1}{\bar{\chi}_1-\chi_1}+\frac{(c_1^2-c_2^2)}{2(\bar{\lambda}_1
     -\lambda_1)}e^{2\mathrm{Re}\left[{\rm i}(\bar{\chi}_1+a_1)[x+\frac{1}{2}(\bar{\chi}_1-a_1)t]\right]},
       \\
     M_2=& \frac{1}{\bar{\chi}_1-\chi_2}e^{{\rm i}\chi_2(x+\frac{1}{2}\chi_2t)},
                    \,\, M_3= \frac{1}{\bar{\chi}_2-\chi_1}e^{-{\rm i}\bar{\chi}_2(x+\frac{1}{2}\bar{\chi}_2t)}
                                                                             ,
                     \\
                     M_4=&\beta+\frac{1}{\bar{\chi}_2-\chi_2}e^{-2\mathrm{Im}(\chi_2)[x+\mathrm{Re}(\chi_2)t]}, \\
   \end{split}
\end{equation*}
and
\begin{equation*}
    \begin{split}
       X_{11}=&   \frac{1}{\chi_1+a_1}+\frac{c_2}{c_1} e^{-{\rm i}(\chi_1+a_1)[x+\frac{1}{2}(\chi_1-a_1)t]}, \,\,
     X_{12}=
                         \frac{1}{\chi_2+a_1}e^{{\rm i}\chi_2(x+\frac{1}{2}\chi_2t)}, \\
      X_{13}=&  e^{-{\rm i}\bar{\chi}_2(x+\frac{1}{2}\bar{\chi}_2t)}\left(\frac{1}{\chi_1+a_1}+
                                                                             \frac{c_2}{c_1} e^{-{\rm i}(\chi_1+a_1)[x+\frac{1}{2}(\chi_1-a_1)t]}\right), \\
      X_{14}=&\frac{1}{\chi_2+a_1}e^{-2\mathrm{Im}(\chi_2)[x+\mathrm{Re}(\chi_2)t]},
    \end{split}
\end{equation*}
and $X_{21}=X_{11}(c_1\leftrightarrow c_2)$, $X_{23}=X_{13}(c_1\leftrightarrow c_2)$, $X_{22}=X_{12}$, $X_{24}=X_{14}$.

\begin{figure}[htb]
\centering
\subfigure[$|q_1|^2$]{\includegraphics[height=50mm,width=80mm]{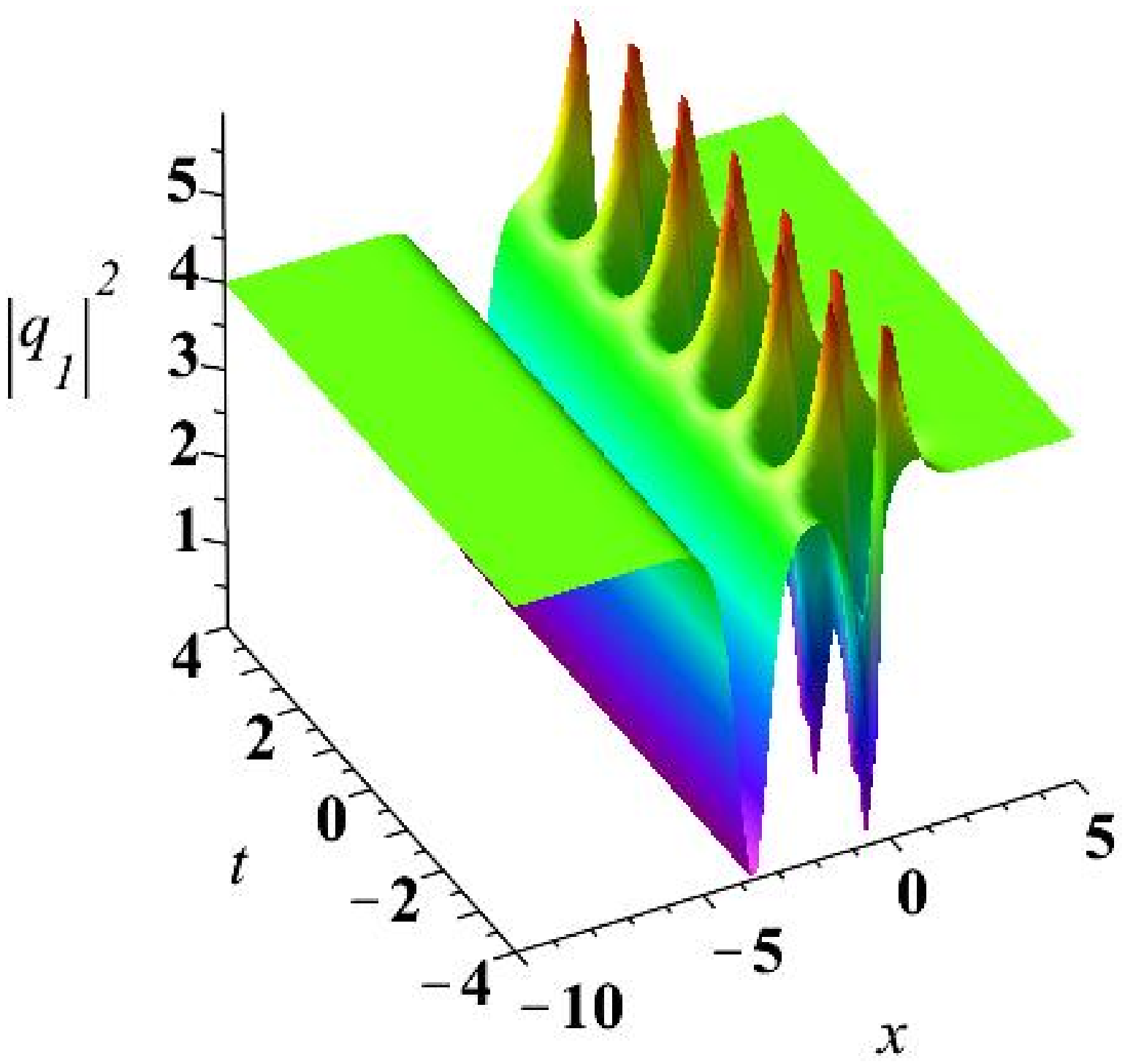}}
\hfil
\subfigure[$|q_2|^2$]{\includegraphics[height=50mm,width=80mm]{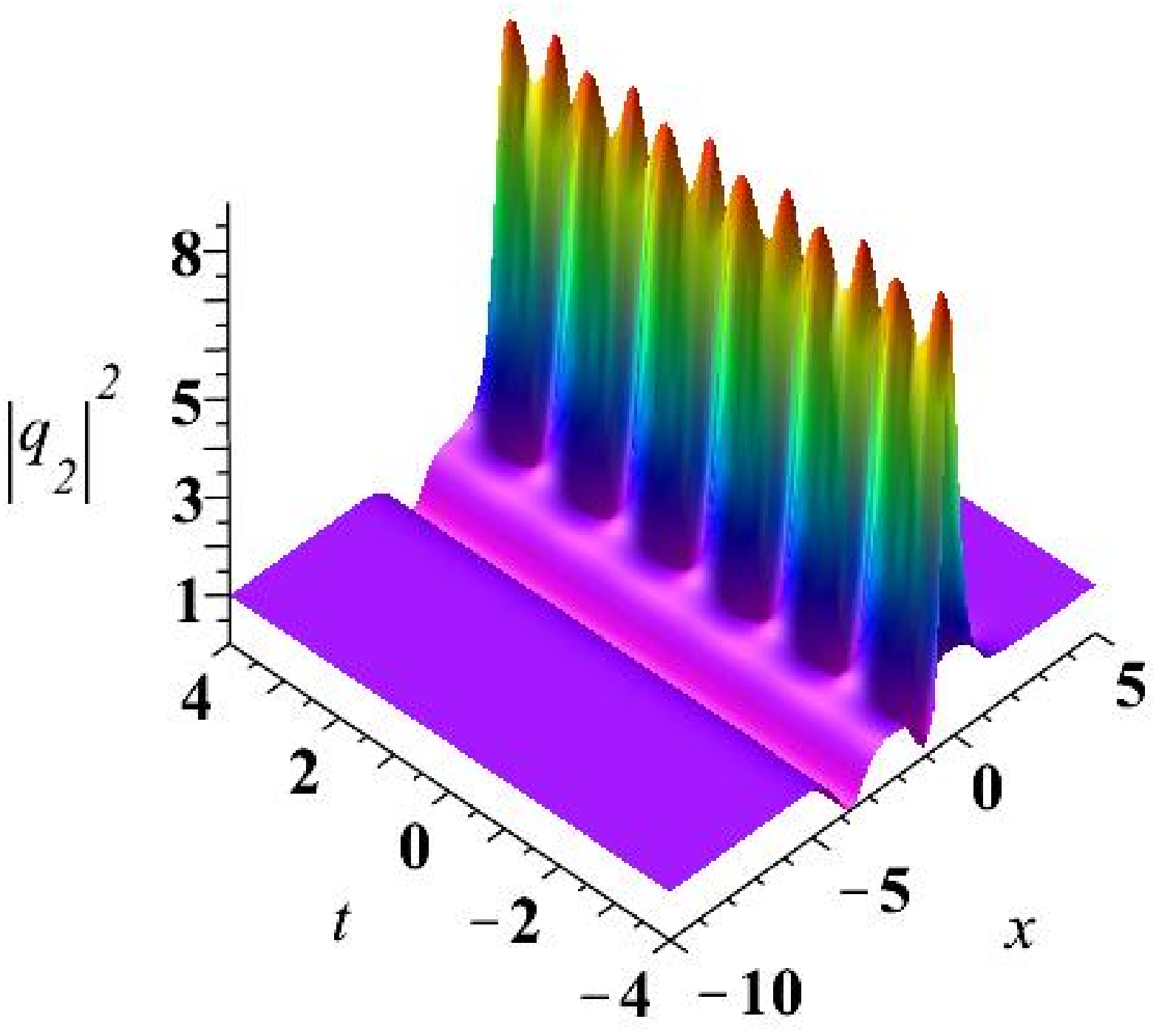}}
\caption{(color online): Dark-dark-breather-II solution: Parameters $a_1=a_2=0$, $c_1=2$, $c_2=1$, $\lambda_1=i$, $\lambda_2
=0$, $\chi_1=3{\rm i}$, $\chi_2=1.732050808{\rm i}$, $\beta=.2886751345\exp(10){\rm i}$. It is seen that
 there exists dark-dark solution and breather-II solution with the stationary.}\label{fig5-1}
\end{figure}

We can perform similar asymptotical analysis on them.  Choosing special parameters, we can obtain
the interaction figure between dark-dark soliton and breather-II solution in the degenerate case.
Here we give an example for the dark-dark soliton and breather-II soliton possesses the same velocity (Fig. \ref{fig5-1}).

\begin{figure}[htb]
\centering
\subfigure[$|q_1|^2$]{\includegraphics[height=50mm,width=80mm]{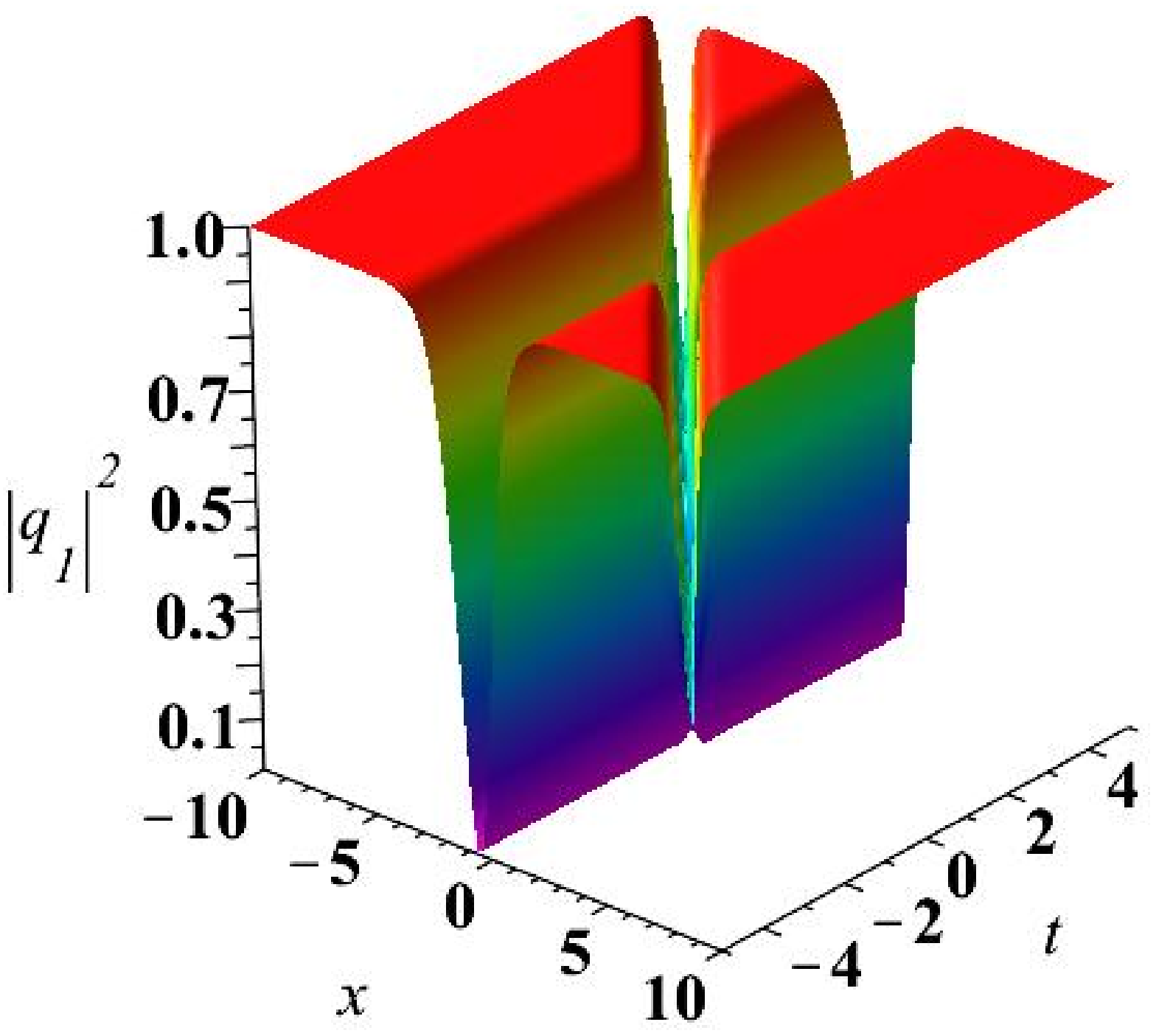}}
\hfil
\subfigure[$|q_2|^2$]{\includegraphics[height=50mm,width=80mm]{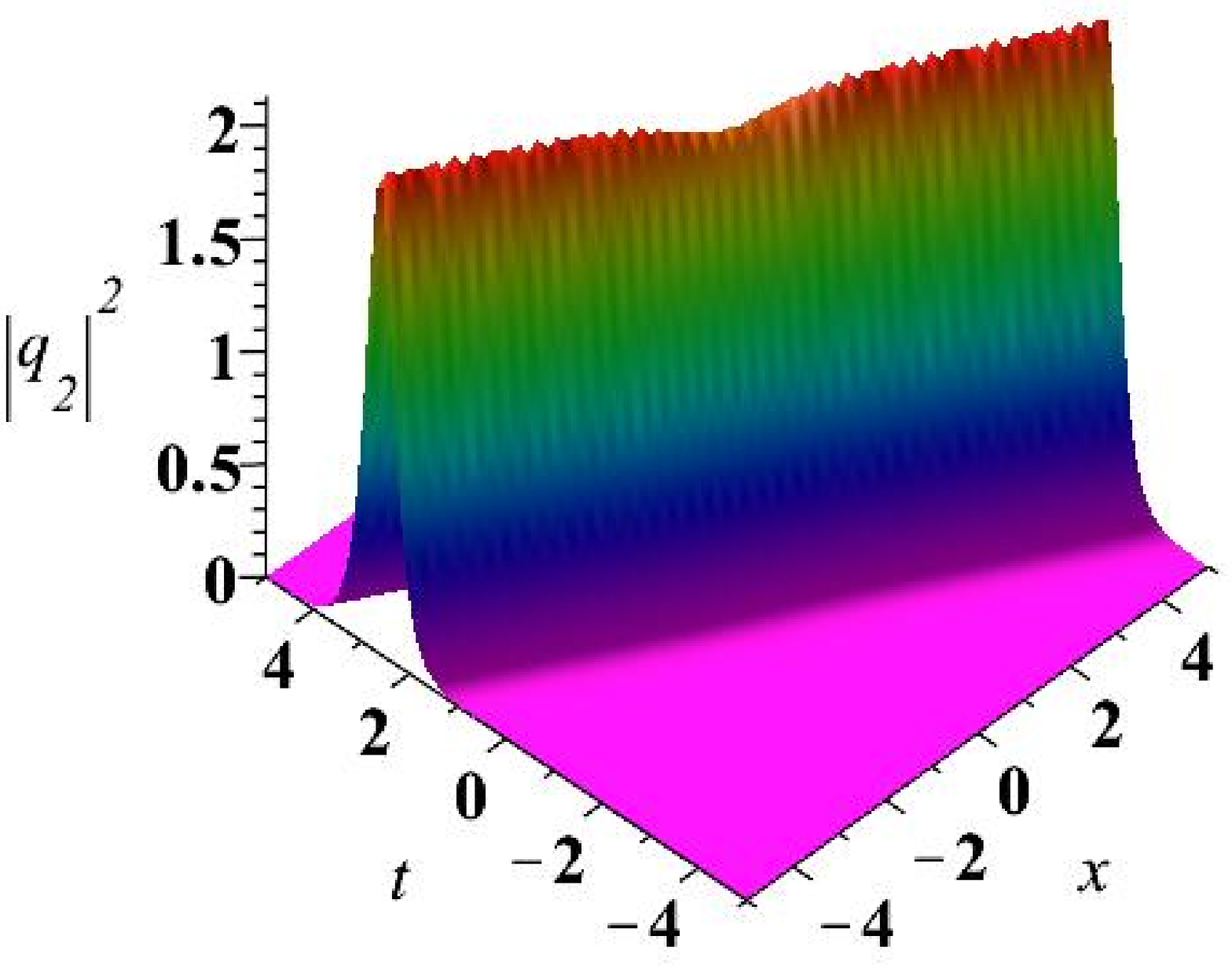}}
\caption{(color online): Two dark-one bright soliton: Parameters $a_1=a_2=0$, $c_1=1$, $c_2=0$, $\lambda_1=1+{\rm i}$, $\lambda_2
=0$, $\chi_1=1.786151378+2.272019650{\rm i}$, $\tau_1=0$, $\chi_2={\rm i}$, $\beta=0.5{\rm i}$. It is seen that there are two dark soliton in the component
$|q_1|^2$ and one bright soliton in the component $|q_2|^2$.}\label{fig7}
\end{figure}

\begin{figure}[htb]
\centering
\subfigure[$|q_1|^2$]{\includegraphics[height=50mm,width=80mm]{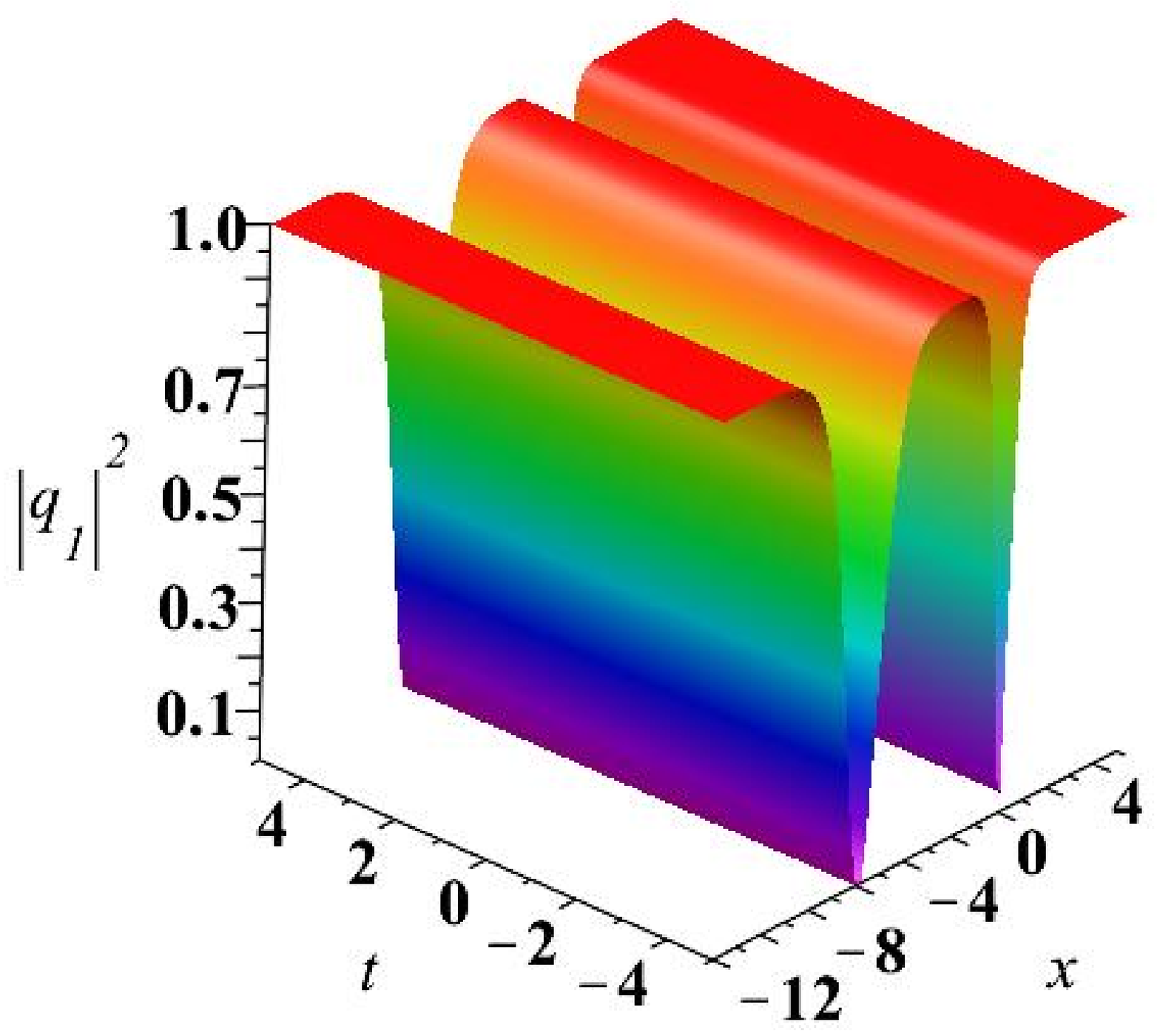}}
\hfil
\subfigure[$|q_2|^2$]{\includegraphics[height=50mm,width=80mm]{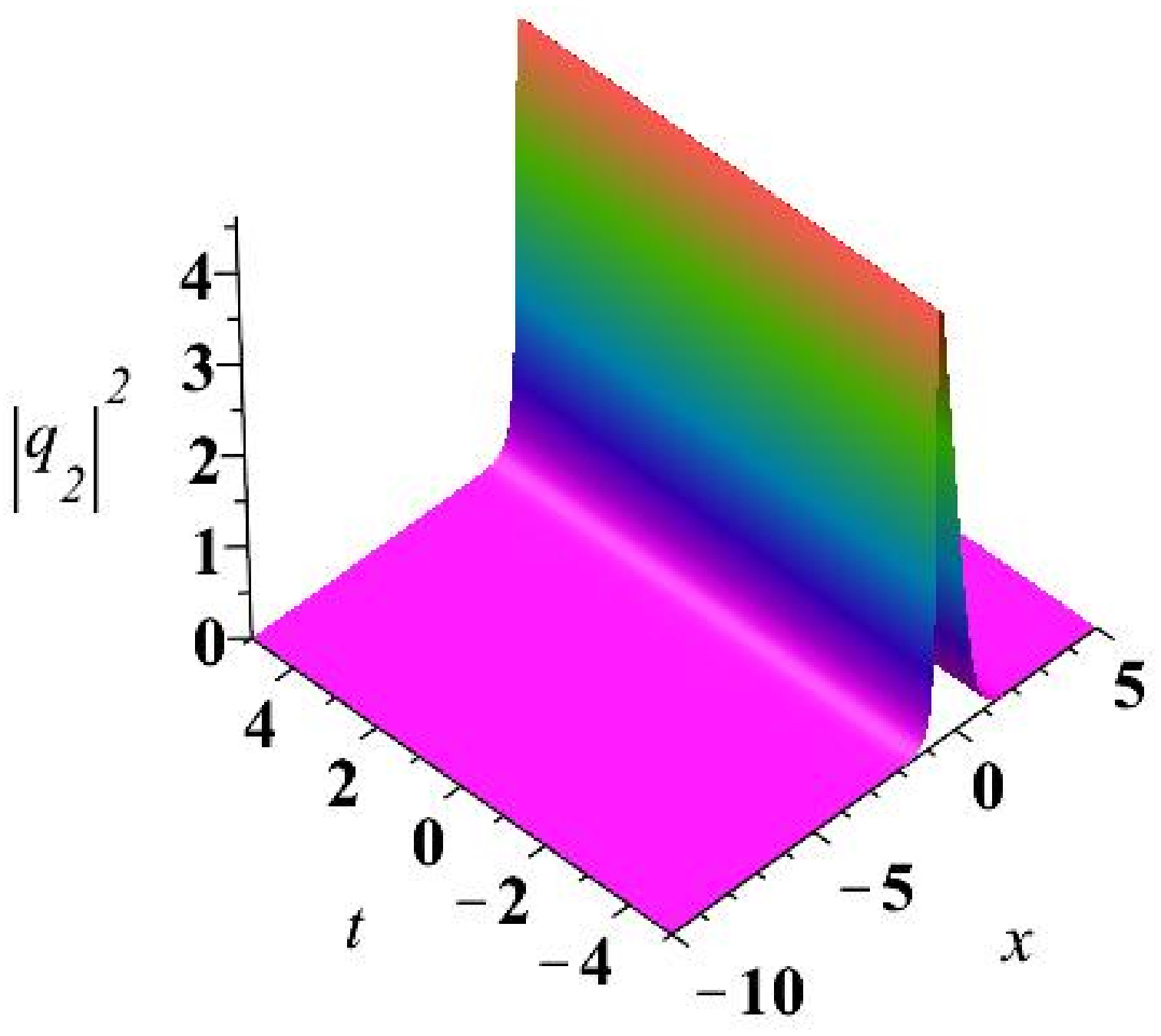}}
\caption{(color online):Two dark-one bright soliton: Parameters $a_1=a_2=0$, $c_1=1$, $c_2=0$, $\lambda_1={\rm i}$, $\lambda_2
=0$, $\chi_1=2.414213562{\rm i}$, $\tau_1=0$, $\chi_2={\rm i}$, $\beta=0.5\exp(10){\rm i}$. It is seen that there are two dark soliton in the component
$|q_1|^2$ with stationary and one bright soliton in the component $|q_2|^2$ with stationary.}\label{fig8}
\end{figure}

Secondly, we consider $c_1>0$ and $c_2=0$. Indeed, the explicit expression of this kind solution can be obtained by
choosing parameter $c_2=0$ on the above. Thus we can have two dark-one bright soliton solution:
\begin{equation}\label{gene-formula1}
    \begin{split}
      q_1[2]=&c_1\frac{\det(M+X_1)}{\det(M)}e^{{\rm i}\theta_1},  \\
      q_2[2]=&\frac{\det(X_2)}{\det(M)}e^{{\rm i}\theta_2},
    \end{split}
\end{equation}
where
\begin{equation*}
      M=\begin{bmatrix}
           M_1 & M_2 \\
           M_3 & M_4 \\
         \end{bmatrix},
        \,\,
       X_1=\begin{bmatrix}
           X_{11} & X_{12} \\
           X_{13} & X_{14} \\
         \end{bmatrix},\,\,X_2=\begin{bmatrix}
           X_{21} & X_{22} \\
           X_{23} & X_{24} \\
         \end{bmatrix}
\end{equation*}
and
\begin{equation*}
   \begin{split}
    M_1=&\frac{1}{\bar{\chi}_1-\chi_1}+\frac{c_1^2}{2(\bar{\lambda}_1
     -\lambda_1)}e^{2\mathrm{Re}\left[{\rm i}(\bar{\chi}_1+a_1)[x+\frac{1}{2}(\bar{\chi}_1-a_1)t]\right]},
       \\
     M_2=& \frac{1}{\bar{\chi}_1-\chi_2}e^{{\rm i}\chi_2(x+\frac{1}{2}\chi_2t)},
                    \,\, M_3= \frac{1}{\bar{\chi}_2-\chi_1}e^{-{\rm i}\bar{\chi}_2(x+\frac{1}{2}\bar{\chi}_2t)}
                                                                             ,
                     \\
                     M_4=&\beta+\frac{1}{\bar{\chi}_2-\chi_2}e^{-2\mathrm{Im}(\chi_2)[x+\mathrm{Re}(\chi_2)t]}, \\
   \end{split}
\end{equation*}
and
\begin{equation*}
    \begin{split}
       X_{11}=&\frac{1}{\chi_1+a_1}, \,\,
     X_{12}=
                         \frac{1}{\chi_2+a_1}e^{{\rm i}\chi_2(x+\frac{1}{2}\chi_2t)}, \\
      X_{13}=&  \frac{1}{\chi_1+a_1}e^{-{\rm i}\bar{\chi}_2(x+\frac{1}{2}\bar{\chi}_2t)},
                                                                              \,\,
      X_{14}=\frac{1}{\chi_2+a_1}e^{-2\mathrm{Im}(\chi_2)[x+\mathrm{Re}(\chi_2)t]}.
    \end{split}
\end{equation*}
and
\begin{equation*}
    \begin{split}
       X_{21}=& c_1e^{-{\rm i}(\chi_1+a_1)[x+\frac{1}{2}(\chi_1-a_1)t]},\,\,
     X_{22}=\left(\frac{1}{\chi_2+a_1}+\frac{1}{\bar{\chi}_1-\chi_2}\right)e^{{\rm i}\chi_2(x+\frac{1}{2}\chi_2t)}, \\
      X_{23}=&  c_1e^{-{\rm i}(\chi_1+a_1)[x+\frac{1}{2}(\chi_1-a_1)t]-{\rm i}\bar{\chi}_2(x+\frac{1}{2}\bar{\chi}_2t)}, \,\,
      X_{24}=\left(\frac{1}{\chi_2+a_1}+\frac{1}{\bar{\chi}_2-\chi_2}\right)e^{-2\mathrm{Im}(\chi_2)[x+\mathrm{Re}(\chi_2)t]}+
      \beta.
    \end{split}
\end{equation*}
To show its dynamics behavior, we choose two different groups of parameters. The figure \ref{fig7} and figure \ref{fig8} show that
the component $|q_1[2]|^2$ possesses two dark soliton and $|q_2[2]|^2$ possesses one bright soliton. In figure \ref{fig7}, the two dark solitons
possess different velocities.  In figure \ref{fig8}, the two dark solitons
possess same velocity with stationary.

\section{Discussions and Conclusions}

In this paper, we provide a method to derive nonsingular nonlinear wave solutions of mixed coupled nonlinear Schr\"{o}dinger equations for which it is essential to deal with indefinite Darboux matrix. Furthermore, we present one possible classification for nonlinear localized wave solutions of the model through combining Darboux transformation and matrix analysis methods. The explicit conditions and ideal excitation forms for these nonlinear waves are presented in detail, which are meaningful for further physical studies on them.
The high order solution can be obtained by the generalized DT \cite{Guo1,He,Guo2,bian}. Indeed, high-order solution can be obtained through limit technique from the solution formula in \textbf{theorem \ref{thm5}} too. We would like to consider the general high order solution with a proper form in the future. The methods here can be extended directly to the defocusing CNLSE or even general multi-component NLSE, the mixed or defocusing Sasa-Satuma system, three wave system, long wave-short wave model and other AKNS reduction system with indefinite Darboux matrix cases.

Recently, a classification on soliton solution for multi-component NLSE was presented in ref. \cite{Tsuchida}, which mainly involving the dark-dark soliton, the bright-dark soliton and the breather solution (called by bright soliton there). The nonlinear wave solutions and the derivation method presented here are distinctive from the results.

It should be pointed out that
the non-singularity condition has been given  through the general algebro-geometric scheme in reference \cite{Dub}. The differences between our work
and reference \cite{Dub} have been given in \cite{Ling-re}.

\section*{Acknowledgments}
This work is supported by National Natural Science Foundation of
China (Contact No. 11401221, 11405129) and Fundamental Research Funds for the Central Universities (Contact No. 2014ZB0034).

\end{document}